\documentclass[preprint,superscriptaddress]{revtex4-2}
\usepackage{amsmath}
\usepackage{bm}
\usepackage{graphicx}
\usepackage[caption=false]{subfig}
\usepackage{amssymb}
\usepackage{booktabs}
\usepackage{makecell}
\usepackage{multirow}
\usepackage{dcolumn}
\usepackage{tikz}
\usepackage[normalem]{ulem}
 
\newcommand*\halfcirc[1][1ex]{%
  \begin{tikzpicture}
  \draw[fill] (0,0)-- (90:#1) arc (90:270:#1) -- cycle ;
  \draw[thick] (0,0) circle (#1);
  \end{tikzpicture}}
\newcommand*\fullcirc[1][1ex]{\tikz\fill (0,0) circle (#1);} 

\begin{document}

\title{Optical Control of Ferroaxial Order}
\date{\today}
\author{Zhiren He}
\email{zhiren.he@unt.edu}
\affiliation{School of Applied and Engineering Physics, \\ Cornell University, Ithaca, New York 14853, USA}
\affiliation{Department of Physics, University of North Texas, Denton, TX 76203, USA}
\author{Guru Khalsa}
\email{guru.khalsa@unt.edu}
\affiliation{Department of Materials Science and Engineering, \\ Cornell University, Ithaca, New York 14853, USA}
\affiliation{Department of Physics, University of North Texas, Denton, TX 76203, USA}
\begin{abstract}
    Materials that exhibit ferroaxial order hold potential for novel multiferroic applications. However, in pure ferroaxials, domains are not directly coupled to stress or static electric field due to their symmetry, limiting the ability to pole and switch between domains -- features required for real-world applications. Here we propose a general approach to selectively condense and switch between ferroaxial domains with light. We show that circularly polarized light pulses on resonance with infrared-active phonons manifest helicity-dependent control over ferroaxial domains. Nonlinear contributions to the lattice polarizability play an essential role in this phenomenon. We illustrate the feasibility of our approach using first-principle calculations and dynamical simulations for the archetypal ferroaxial material RbFe(MoO$_4$)$_2$. Our results are discussed in the context of future pump-probe optical experiments, where polarization, carrier frequency, and fluence threshold are explored.
\end{abstract}
\maketitle
\section{Introduction \label{sec:intro}}
Materials functionalities derive from the changes in materials properties induced by external fields. In ferroelectrics, switching of the electric polarization through an applied voltage (or electric field) is leveraged for electrical control of capacitance or metastable ferroelectric \emph{bits} of memory \cite{Devonshire_1954}.  In an intimately connected phenomenon piezoelectricity, applied external stress is transformed into an electric field which can be read as the voltage change across a capacitor. Similarly, ferromagnetic domains may be switched by the application of a magnetic field. However, other functionalities may be hidden due to the challenge of finding an external field that couples to the desired property \cite{Hayami_2018}. One such class of materials termed \emph{ferroaxial} or ferro-rotational materials \cite{Gopalan_2011, Cheong_2018, Hlinka_2016}, have been explored for their potential multiferroic properties, enabling control of magnetism through an electric bias \cite{White_2013, Hearmon_2012, Zhang_2011, Johnson_2011, Johnson_2012, Li_2012, Cao_2014}.

The challenges and opportunities in ferroaxial materials derive from the symmetry of the ferroaxial order parameter $\bm{A}$, which is composed of local electric dipole moments $\bm{d}$ oriented in closed loops ($\bm{A} \propto \bm{r}\times\bm{d}$). On one hand, once condensed, $\bm{A}$ allows for direct coupling between magnetism and the electric polarization \cite{Cao_2014}. On the other hand, direct coupling of a ferroaxial mode with a static electric field or even stress may be forbidden \cite{Hlinka_2016}, therefore poling, switching, and characterizing ferroaxial domains pose substantial experimental hurdles \cite{Hayashida_2020,Hayashida_2021,Fang_2023}.

The development of intense and ultrafast mid- and far-infrared light sources provides a new perspective on structural control of materials and access to functional properties \cite{Subedi_2014, Radaelli_2018, Disa_2021}. In the conventional nonlinear phononics approach, resonant excitation of infrared(IR)-active phonons $Q_{\text{IR}}$ to large amplitude via short-duration pulses ($<$ 1 ps) transiently induces structural changes through coupled Raman-active phonons $Q_{\text{R}}$. This is accomplished by large anharmonic lattice coupling in the energy $\propto Q_{\text{R}}Q_{\text{IR}}^2$ \cite{Mankowsky_2014}. Recent theoretical work suggests that phonons of arbitrary symmetry $Q_{\alpha}$, not only Raman-active phonons, may dominate the structural changes via higher-order anharmonic lattice energy $ \propto Q_{\alpha}^2Q_{\text{IR}}^2$ \cite{khalsa_2023}, and that nonlinear contributions to the lattice polarizability should be held on equal footing with the anharmonic lattice energy contributions to the transient structural response \cite{Khalsa_2021,Blank_2023}. 

\begin{figure*}[!ht]
    \centering
    \includegraphics[width=0.75\textwidth]{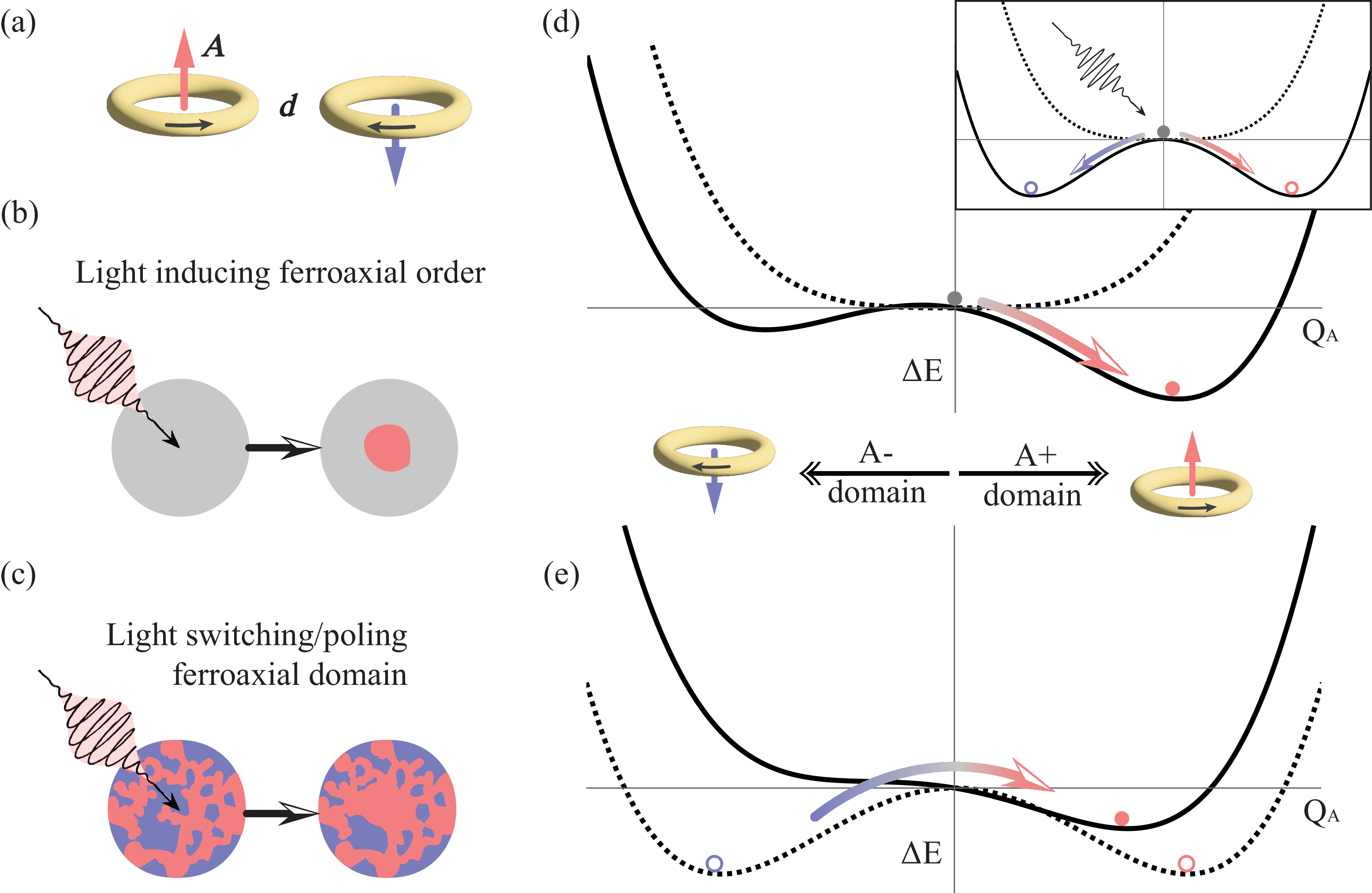}
    \caption{\label{fig:main} \textbf{Mechanism for light-controlled ferroaxial order}. (a) Ferroaxial order is composed of electric dipolar loops schematically represented by a pseudovector $\bm{A}$ pointing up (red) or down (blue). (b) Circularly polarized light, resonant with IR-active phonons selectively and transiently induces a ferroaxial domain from the para-axial phase; or, in (c) the polydomain ferroaxial phase, this excitation scheme enables switching between ferroaxial domains. (d) Schematic of the light-induced phase transition upon IR-active phonon excitation. Circularly polarized light softens and biases the equilibrium potential (dotted line) to a tilted double-well potential (solid line), selecting a single domain via the light helicity. Inset: linearly polarized light can induce a double well, but cannot bias the ferroaxial order. (e) Schematic of the light-enabled domain control. Circularly polarized light stiffens and biases the double-well potential (dotted line) towards a tilted single well (solid line), poling the domain, whose direction is determined by light helicity. }
\end{figure*}

Here we show that the nonlinear lattice polarizability (NLP) gives a natural and general handle on the ferroaxial mode, suggesting that transient structural control of ferroaxial materials is possible via the nonlinear phononics mechanism. We flesh out this perspective by constructing the phenomenological free energy and then perform density functional theory (DFT) calculations to confirm the feasibility of this approach in an archetypal ferroaxial material RbFe(MoO$_4$)$_2$, whose ferroaxial mode has no simple coupling to electric field or stress. By computing the NLP and anharmonic lattice energy contributions, we explore the energetics and dynamics of the transient structural changes in RbFe(MoO$_4$)$_2$, finding the condensation and control of the ferroaxial order to be within the experimental capability of existing mid-IR laser sources. An overview of the proposed approach to ferroaxial control is shown in FIG.~\ref{fig:main}.

The organization of the paper is as follows: We start by developing a simple, but general model for the response of the ferroaxial mode to excitation of an IR-active phonon. We then explore the dynamics and threshold phenomenon of the ferroaxial response guided by parameters taken from first-principles calculations, first showing light-induced transient single-domain ferroaxial order, then deterministic ferroaxial domain switching with light. Finally, we discuss possible experimental strategies to confirm our predictions followed by a summary and outlook.

\begingroup
\squeezetable
\begin{table*}[!ht]
\begin{ruledtabular}
    \begin{tabular}{ccccccc}
        No. & G & \makecell{Ferroaxial \\ irrep.} &\makecell{NLP  $\left( \bm{E}^*\times \bm{Q}_\text{IR} \right)
\cdot \bm{Q}_\text{A}$ }& F & Ferroaxial & Index   \\ \midrule
        1 & 4mm (C$_{4v}$) & A$_2$  & \{E $\boxtimes$ E\}A$_2$ & 4 (C$_4$) & \fullcirc & 2  \\ \midrule
        2 & $\bar{4}$2m (D$_{2d}$) & A$_2$ & \{E $\boxtimes$ E\}A$_2$ & $\bar{4}$ (S$_4$) & \fullcirc & 2   \\ \midrule
        3 & 4/mmm (D$_{4h}$) & \multirow{2}{*}{A$_{2g}$} & \multirow{2}{*}{\{E$_u$ $\boxtimes$ E$_u$\}A$_{2g}$}& 4/m (C$_{4h}$) & \fullcirc & 2   \\ 
        4 & 4/mmm (D$_{4h}$) & ~ &  ~ & $\bar{4}$ (S$_4$) & \halfcirc & 4 \\ \midrule
        5 & 3m (C$_{3v}$) & A$_2$ & \{E $\boxtimes$ E\}A$_2$ & 3 (C$_3$) & \fullcirc & 2    \\ \midrule
        6 & $\bar{3}$m (D$_{3d}$) & A$_{2g}$ &  \{E$_u$ $\boxtimes$ E$_u$\}A$_{2g}$& $\bar{3}$ (C$_{3i}$) & \fullcirc & 2  \\ \midrule
        7 & 6mm (C$_{6v}$) & \multirow{2}{*}{A$_2$} & \multirow{2}{*}{\{E$_1$ $\boxtimes$ E$_1$\}A$_2$ }& 6 (C$_6$) & \fullcirc & 2   \\ 
        8 & 6mm (C$_{6v}$) & ~ & ~& 3 (C$_3$) & \halfcirc & 4   \\ \midrule
        9 & $\bar{6}$m2 (D$_{3h}$) & A$_2'$ &  \{E$'$ $\boxtimes$ E$'$\}A$_2'$& $\bar{6}$ (C$_{3h}$) & \fullcirc & 2  \\ \midrule 
        10 & 6/mmm (D$_{6h}$) & \multirow{3}{*}{A$_{2g}$} & \multirow{3}{*}{ \{E$_{1u}$ $\boxtimes$ E$_{1u}$\}A$_{2g}$}& 6/m (C$_{6h}$) & \fullcirc & 2  \\ 
        11 & 6/mmm (D$_{6h}$) & ~ &  ~ & $\bar{6}$ (C$_{3h}$) & \halfcirc & 4 \\ 
        12 & 6/mmm (D$_{6h}$) & ~ & ~& $\bar{3}$ (C$_{3i}$) & \halfcirc & 4 \\ 
    \end{tabular}
    \end{ruledtabular}
    \caption{\label{tab:Aizu} The 12 \emph{pure ferroaxial} species \cite{Hlinka_2016}. G and F are the point groups of the high- and low-symmetry structures related by the ferroaxial irreducible representation. The ferroaxial irreducible representation can be coupled via an antisymmetric product (\{...$\boxtimes$...\}) of 2D irreducible representations that transform as vectors (i.e. \{$x,y$\}). The forms of the NLP for the pure ferroaxial species are shown in the fourth column. Filled circles represent full ferroaxiality where the ferroaxial mode is the sole order-parameter. Half-filled circles represent partial ferroaxiality where the ferroaxial mode works in concert with an additional mode to define the symmetry in F. The index gives the factor of symmetry elements lost from G in the transition to F (i.e. $|G|/|F|$).}
\end{table*}
\endgroup

\section{Phenomenology of ferroaxial mode coupling \label{sec:phenomenology}}
Four types of fundamental multipole order parameters, classified according to their spatial inversion and time-reversal properties, define a complete basis set for describing coupled charge, spin, and orbital orders and their coupling to external fields  \cite{Hayami_Micro, Hayami_2018}. The electric (magnetic) dipole moment, a vector order parameter for the charge (spin), is odd under inversion I (time-reversal TR) and therefore couples directly to an externally applied electric (magnetic) field. The magnetic toroidal dipole breaks both I and TR and therefore couples to simultaneously applied magnetic and electric fields \cite{Spaldin_2008, Zimmermann_2014}. The fourth fundamental vector order parameter is the electric toroidal dipole or ferroaxial order, which is even under both I and TR. How does ferroaxial order couple to external fields? In some crystal systems, the ferroaxial mode is indistinguishable by symmetry from an electric polarization or strain. In such cases, an electric bias or stress may be a viable option for accessing the ferroaxial mode. In a recent experimental demonstration, Yuan et al. switched ferroaxial domains through the application of uniaxial stress in CaMn$_7$O$_{12}$ \cite{Yuan_2015}. But this is not always the case. In so-called \emph{pure ferroaxials} (TABLE~\ref{tab:Aizu}), neither electric field nor stress can affect the ferroaxial mode, and no simple method for its control is known. What then is the lowest-order coupling to a ferroaxial mode?

The ferroaxial order $\bm{A} \propto \sum_i \bm{r}_i\times\bm{d}_i$ is an anti-symmetric product of two vectors representing the positions $\bm{r}_i$ of the dipoles, and their dipole moments $\bm{d}_i$. Its conjugate field must therefore transform like the anti-symmetric product of two vector fields, which vanishes in most cases. Without loss of generality, consider a pure ferroaxial, whose orientation pseudovector is aligned with the $z$-axis, that we intended to control with static electric fields oriented in the $xy$-plane. The antisymmetric product of the electric field is $\left\{\bm{E} \times \bm{E}\right\}\cdot \hat{z} = E_x E_y - E_y E_x = 0$, due to the commutation of fields. This explains the difficulty of altering a pure ferroaxial by means of an electric field or, following the same argument, stress alone. Though a combination of electric field and stress can resolve this issue through higher-order couplings \cite{Hlinka_2016}, the effective field is expected to be weak, and to the best of our knowledge, no such experimental demonstration exists.

We propose an alternative solution to this problem by utilizing two dynamical vector fields: a light field $\bm{E}$, and an IR-active phonon $\bm{Q}_{\text{IR}}$. In contrast to the static case described above, light can be prepared with circular or elliptical polarization so that $\bm{E}^*\times \bm{Q}_\text{IR}\ne0$. Notice that $ \bm{E}^*\times \bm{Q}_{\text{IR}}$ is an axial vector and therefore couples directly to the ferroaxial mode $\bm{Q}_{\text{A}}$. This coupling represents a nonlinear part of the polarizability of the crystal involving two phonons and one photon, which contribute to the energy as 
\begin{equation} \label{eqn:NLP}
\begin{split}
\Delta \mathcal{F}=& \mathfrak{Re}\left\{\tilde{C_c} \left[ \bm{E}^*(\omega_1)\times \bm{Q}_{\text{IR}}(\omega_2) \right]
\cdot \bm{Q}_\text{A}(\omega_3)\right\}\\ = &\mathfrak{Re}\Bigl\{\tilde{C_c} i\bigl[ \bm{E}_L^*(\omega_1) \bm{Q}_{\text{IR}, L}(\omega_2) \\ - & \bm{E}_R^*(\omega_1) \bm{Q}_{\text{IR}, R}(\omega_2)\bigl]
\cdot \bm{Q}_\text{A}(\omega_3)\Bigl\} \\ \equiv &\Delta \bm{P}_{\text{NL}} \cdot \bm{E} 
\end{split} 
\end{equation}
where $\omega_3=\omega_1-\omega_2$ for energy conservation. Here,  $\bm{E}(\omega)$ is the Fourier coefficient of the electric field at frequency $\omega$, with $\bm{E}^*(\omega) = \bm{E}(-\omega)$, its complex conjugate, and similarly for $\bm{Q}_\text{IR}(\omega)$. We denote $\tilde{C_c}$ as the complex nonlinear lattice polarizability coupling strength. More explicitly, the nonlinear lattice polarization $\Delta P_{\text{NL},i}=\tilde{C_c}\varepsilon_{ijk}Q_{\text{IR},j}Q_{\text{A},k}$, where $\varepsilon_{ijk}$ is the Levi-Cevita symbol. This form of the free energy is ubiquitous in all 32 point groups, and therefore all 230 space groups representing crystals in three dimensions. As a result, ferroaxial modes leading to pure ferroaxial phase transitions can be directly accessed through circularly (elliptically) polarized light, as detailed in TABLE~\ref{tab:Aizu}.

Eqn.~\ref{eqn:NLP} suggests that the maximum effect of the nonlinear polarizability on the ferroaxial mode is acheived through resonant excitation of $\bm{Q}_{\text{IR}}$ to large amplitude. We illustrate how this gives a general handle on $\bm{Q}_{\text{A}}$ by considering a resonant excitation of an IR-active phonon with circularly polarized light, using a continuous-wave excitation to demonstrate the basic physical result. In this case, the electric field and induced $\bm{Q}_{\text{IR}}$ take the form
\begin{equation} \label{eqn:plane-wave_simple}
\begin{array}{l}
    \mathfrak{Re}\left\{\bm{E}_{\text{L/R}}\right\} \\=  \mathfrak{Re}\left\{ \frac{E_0}{2} \begin{bmatrix}
       1 \\  \pm i
    \end{bmatrix}e^{-i\omega_0t}\right\} =  \frac{E_0}{\sqrt{2}} \begin{bmatrix}
       \cos(\omega_{0}t) \\  \pm \sin(\omega_{0}t)
    \end{bmatrix}\\
    \mathfrak{Re}\left\{\bm{Q}_{\text{IR, L/R}}\right\} \\=  \mathfrak{Re} \left\{-i\frac{Q_0}{2} \begin{bmatrix}
       1 \\  \pm i
    \end{bmatrix}e^{-i\omega_0t}\right\} =  \frac{Q_0}{\sqrt{2}} \begin{bmatrix}
       -\sin(\omega_{0}t) \\  \pm \cos(\omega_{0}t)
    \end{bmatrix} 
\end{array}
\end{equation}
where $E_0$ and $Q_0$ are the amplitudes of the electric field and IR-active phonon, and $\omega_{0}$ is the resonant frequency of the IR-active phonon, in the limit of $\omega_1\approx\omega_2=\omega_0$, and $\omega_3\rightarrow0$ in Eqn. \ref{eqn:NLP}. The $+/-$ is chosen for left/right-circularly polarized light. The phase difference of $\pi/2$ between $\bm{E}$ and $\bm{Q}_{\text{IR}}$ represented as $-i$ in the second equation is due to \emph{phase lag}, a phenomenon seen in all harmonic oscillators driven on resonance, such that the composite object $\bm{E}^*(\omega_1)\times \bm{Q}_{\text{IR}}(\omega_2)$ is even under TR. Inserting Eqn.~\ref{eqn:plane-wave_simple} into Eqn.~\ref{eqn:NLP} gives
\begin{equation}
	\begin{split}
		&\left(\Delta\bm{P}_{\text{NL}}\cdot\bm{E}\right)_{\text{L/R}}\\ = & \frac{1}{4}Q_{\text{A},z}E_{0}Q_{0} Re\left\{ -i  \tilde{C}_c\begin{bmatrix}
			1 \\  \mp i
		\end{bmatrix}\times \begin{bmatrix} 
			1 \\  \pm i
		\end{bmatrix}\right\} \\ = & \pm\frac{1}{2} C_cQ_{\text{A},z}E_{0}Q_{0}
	\end{split}
\end{equation}
where only $\mathfrak{Re}\{\tilde{C}_c\}=C_c$ is contributing due to the $\pi/2$ phase lag of $\bm{Q}_{\text{IR}}$. This shows that resonant excitation of an IR-active phonon with circularly polarized light gives a direct strategy for directional control of the ferroaxial mode through the helicity of the light, with left-circularly polarized light biasing the ferroaxial order in one direction, and right-circularly polarized light biasing the order in the other direction (FIG.~\ref{fig:main}). We highlight here that the unidirectional drive on the ferroaxial mode from this term is expected to be (quasi)static since no oscillatory component from the light field or phonon appears in the equation. In Appendix~\ref{app:NLP} we show that both near-resonant excitation of the IR-active phonon and use of elliptical polarization are necessary for this result in the continuous wave and impulsive excitation limits.

We now formulate an expansion of the free energy of the lattice system in the presence of an electric field. We focus on the average coordinates $\bm{Q}_{\text{IR}}$ and $\bm{Q}_{\text{A}}$, and again let $\bm{Q}_{\text{A}}$ be along the $z$-axis and $Q_{\text{A},z}\equiv Q_{\text{A}}$. Expanding the lattice energy to fourth order in the phonon coordinates and including the coupling to an electric field we find:
\begin{equation} \label{eqn:energy}
	\begin{split}
    \mathcal{F} =& \frac{1}{2}K_{\text{IR}} \bm{Q}_{\text{IR}}^2 
    + \frac{1}{2}K_{\text{A}}Q_{\text{A}}^2 
    + \frac{1}{4} D_{\text{A}}Q_{\text{A}}^4 \\
    + & D_{22} \bm{Q}_{\text{IR}}^2Q_{\text{A}}^2
    - \left(\tilde{Z}^*\bm{Q}_{\text{IR}}+\Delta\bm{P}_{\text{NL}}\right)\cdot\bm{E}
    \end{split}
\end{equation}
Here $K_{\text{IR/A}}$ is the second-order force constant of the IR-active/ferroaxial mode where $\bm{Q}_{\text{IR}}^2 = \bm{Q}_{\text{IR}}^\top\bm{Q}_{\text{IR}} = Q_{\text{IR},x}^2 + Q_{\text{IR},y}^2$. $D_{\text{A}}$ is the fourth-order force constant of the ferroaxial mode. The lowest-order anharmonic coupling between $\bm{Q}_{\text{IR}}$ and $Q_{\text{A}}$ is biquadratic because the linear-quadratic couplings, e.g. $Q_{\text{A}}\bm{Q}_{\text{IR}}^2$ and $Q_{\text{A}}^2\bm{Q}_{\text{IR}}$, are not allowed for the ferroaxial mode. Notice that if $D_{22} < 0$, and the IR-active phonon is displaced to large amplitude, we can collect terms to define an effective force constant of $Q_{\text{A}}$ as $K^*_{\text{A}} = \left(K_{\text{A}} -2\vert D_{22}\vert \bm{Q}_{\text{IR}}^2  \right)$ which may become negative. Thus $Q_{\text{A}}$ may be condensed dynamically via the excitation of $\bm{Q}_{\text{IR}}$ \cite{Khalsa_2021}. In the last term, $\tilde{Z}^*$ is the mode-effective charge of the IR-active phonon. 

The coherent dynamical response of $Q_\text{A}$ driven by infrared light through $\bm{Q}_{\text{IR}}$ depends on the parameters in the model derived above. We now look at an archetypal ferroaxial material, RbFe(MoO$_4$)$_2$, with parameters drawn from first-principles calculations.

\section{Simulated response of R\lowercase{b}F\lowercase{e}(M\lowercase{o}O$_4$)$_2$ \label{sec:RFMO}}
\subsection{Overview\label{sec:overview}} 
\begin{figure}[!ht]
	\centering
\includegraphics[width=0.5\textwidth]{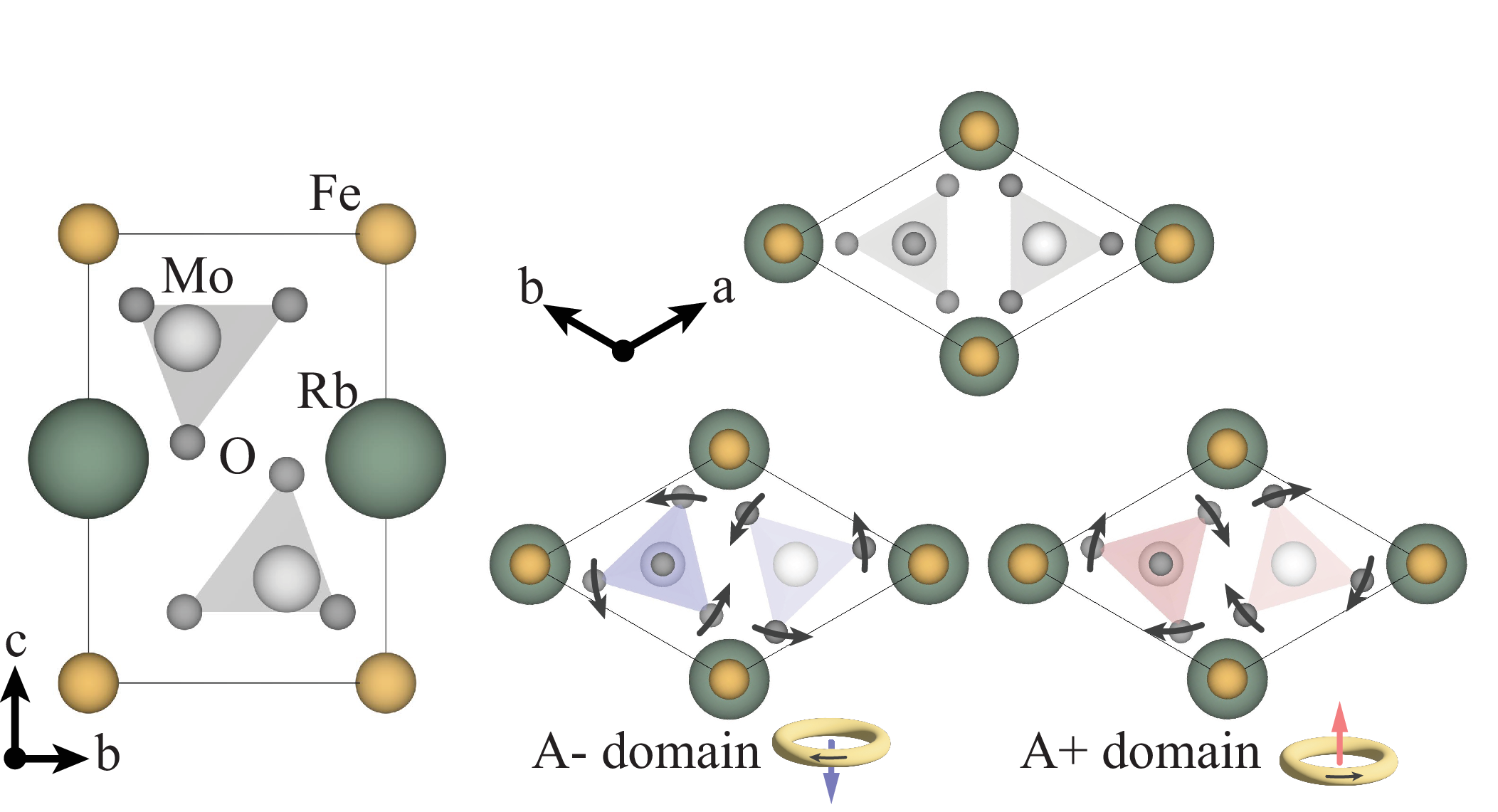}
	\caption{An archetypal ferroaxial, RbFe(MoO$_4$)$_2$ viewed along the $a$ and $c$ crystallographic axes in the paraaxial phase. Right bottom: in the ferroaxial phase, the electric dipolar loops are initiated by coherent rotations of the Mo-O tetrahedra, establishing the two ferroaxial domains.}
	\label{fig:struct}
\end{figure}

RbFe(MoO$_4$)$_2$ belongs to the trigonal double-molybdates/tungstates crystal family,  characterized by its layered structure of alternating Rb$^{1+}$ and Fe$^{3+}$ layers, separated by MoO$_4^{2-}$ tetrahedrons \cite{Klevtsov1977, Zapart2021} (FIG.~\ref{fig:struct}). At 195 K, a ferroaxial phase transition involving primarily rotation of MoO$_4^{2-}$ tetrahedrons along the $c$-axis takes place. The appearance of the ferroaxial phase is represented by a primary order parameter transforming as a 1-dimensional $\Gamma_{2+}$ irreducible representation which reduces the symmetry from P$\bar{3}$m1 to P$\bar{3}$ by removing the two-fold rotational axis and mirror-planes of the parent space group. Two opposite domains with clockwise/counter-clockwise rotation of the MoO$_4^{2-}$ tetrahedra (FIG.~\ref{fig:struct}), having characteristic sizes from 10 $\mu$m - 100 $\mu$m occur in experiments, with differing domain patterns observed upon thermal cycling \cite{Jin_2019, Hayashida_2021}. A sudden change in the $a$ and $c$ lattice constants accompanies the ferroaxial phase transition, which has been described as weakly first-order \cite{Wakowska_2010, Jin_2019, Hayashida_2021}. The multiferroic potential of RbFe(MoO$_4$)$_2$ emerges at 3.8 K where it undergoes a magnetic phase transition from the paramagnetic state to a 120$^\circ$ in-plane antiferromagnetic and incommensurate out-of-plane magnetic configuration \cite{Inami_2007}.
\subsection{Model parameters \label{sec:params}} 
In our exploration of light-induced ferroaxial order, we have found that the doubly-degenerate IR-active mode No. 9 at 23.7 THz, visualized in SI III, has the necessary ingredients for a large response: (1) the highest mode-effective charge $\tilde{Z}^*$, which signals the efficient transfer of intensity from the exciting optical field to phonon amplitude; (2) the largest negative biquadratic coefficient $D_{22}$, which will strongly soften the effective force constant of the ferroaxial mode; and (3) a large nonlinear polarizability coefficient $C_c$, which enables the unidirectional control with circularly polarized light. However, the large negative $D_{22}$ makes the 23.7 THz mode a poor choice for domain switching. Instead, we demonstrate the switching process using an 8.6 THz IR-active mode, where $D_{22}$ is slightly positive, in Section~\ref{sec:switch}.

We extract the model parameters $K_{\text{IR}}$, $D_{22}$, $K_{\text{A}}$ and $D_{\text{A}}$ for RbFe(MoO$_4$)$_2$ from DFT calculations by fitting energy surfaces. $C_c$ was obtained by fitting the change in IR-active mode effective charge with respect to the ferroaxial mode amplitude. The parameters are summarized in TABLE~\ref{tab:model}, and computational details can be found in Appendix~\ref{app:dft} and SI I.
\begingroup
\squeezetable
\begin{table}
\begin{ruledtabular}
\begin{tabular}{ccccc}
    Coeff & \makecell{IR No. 9 \\ 23.7 THz} & \makecell{IR No. 6 \\ 8.6 THz} & Coeff & \makecell{Ferroaxial \\mode}\\ \cmidrule(lr){1-1}\cmidrule(lr){2-3}\cmidrule(lr){4-4}\cmidrule(lr){5-5}
    $\tilde{Z}^*$ (e) & 8.61 & 1.59 & $\tilde{Z}^*$ (e) & 0\\ 
    $M_\text{IR}$ (AMU) & 19.15 & 16.65 & $M_\text{A}$ (AMU)& 16\\
    $K_{\text{IR}}$ (meV/\AA$^2$) & 4.438 $\times 10^4$ & 5.10 $\times 10^3$ & $K_{\text{A}}$ (meV/\AA$^2$) & 28.7\\
    $D_{22}$ (meV/\AA$^4$) & -7676  & 917 & $D_{\text{A}}$ (meV/\AA$^4$) & 1271\\
    $C_c$ (e/\AA)& 0.638  & 0.450 & & 
\end{tabular}
\end{ruledtabular}
\caption{A summary of model parameters used in dynamical simulation. All parameters are obtained from DFT calculations or fitting of DFT results.} \label{tab:model}
\end{table}
\endgroup

\subsection{Dynamical simulation methods\label{sec:simulation_method}} 

Equating the derivative of Eqn.~\ref{eqn:energy} with respect to the phonon coordinates to the time-rate-of change of the momentum of the phonon coordinates and including a phenomenological damping term leads to the following equations of motion for the IR-active and ferroaxial phonons
\begin{widetext}
\begin{equation} \label{eqn:EOM}
\left\{
\begin{array}{l}
	M_{\text{IR}} \ddot{\bm{Q}}_{\text{IR}} + 2 M_{\text{IR}} \gamma_{\text{IR}}\dot{\bm{Q}}_{\text{IR}} + (K_{\text{IR}}+2D_{22}Q_{\text{A}}^2)\bm{Q}_{\text{IR}} = \begin{bmatrix}
		Z^* & -C_cQ_{\text{A}} \\ C_cQ_{\text{A}} & Z^*
	\end{bmatrix}\bm{E}(t)\\
	M_{\text{A}} \ddot{Q}_{\text{A}} + 2 M_{\text{A}} \gamma_{\text{A}}\dot{Q}_{\text{A}} + (K_{\text{A}}+2D_{22}\bm{Q}_{\text{IR}}^\top \bm{Q}_{\text{IR}})Q_{\text{A}} + D_{\text{A}}Q_{\text{A}}^3 = \bm{Q}_{\text{IR}}^\top \begin{bmatrix}
		0 & -C_c \\ C_c & 0
	\end{bmatrix}\bm{E}(t)
\end{array} \right.
\end{equation}
\end{widetext}
These equations of motion can be numerically solved to find the coherent response of the phonon degrees of freedom to a mid-infrared pulse with parameters for RbFe(MoO$_4$)$_2$ taken from first-principles calculations. $M_{\text{IR}}$ and $M_{\text{A}}$ are the mode-effective masses (TABLE~\ref{tab:model}). The damping parameter $\gamma_{\text{IR}}$ $\approx$ 4 THz is estimated from the experimental absorbance data, presented in SI III, and
$\gamma_{\text{A}}$ is assumed to be 1 THz \footnote{A damping term of the form $\gamma_\text{L}\left(\bm{Q}_\text{IR}\times\bm{\dot{Q}}_\text{IR}\right)\cdot\bm{Q}_\text{A}$ which couples the angular momentum of the IR-active phonon to the ferroaxial mode may also contribute to the ferroaxial dynamics. The first principle study of $\gamma_\text{L}$ is extremely cumbersome with current computational resources.}.

We use a Gaussian pulse to describe the electric field:
\begin{equation}
	\bm{E}(t)=
    \frac{E_0}{\sqrt{2}}\begin{bmatrix}
        \cos{\left(\omega t \right)} \\ \cos{\left(\omega t + \phi\right)}
    \end{bmatrix}
    \exp\left\{-\frac{(t-t_0)^2}{2\sigma^2}\right\}. 
\end{equation}
$\phi=0,\ \pi/2,\ -\pi/2$ correspond to linear, left-circular, and right-circular polarization. $\sigma$ adjusts the pulse duration and is related to the full-width half-maximum of the electric field by $\tau_{\text{FWHM}} = 2\sigma\sqrt{2\text{ln}2}$. $E_0$ is the peak electric field. Notice that with linearly polarized light, terms with $C_c$ fall out of Eqn.~\ref{eqn:EOM}, leaving only terms with $D_{22}$ to contribute to the coupling between the driven $\bm{Q}_{\text{IR}}$ and $Q_{\text{A}}$.

\subsection{Excitation of single-domain ferroaxial order \label{sec:simulation}}
\begin{figure*}[!ht]
	\centering
	\includegraphics[width=\textwidth]{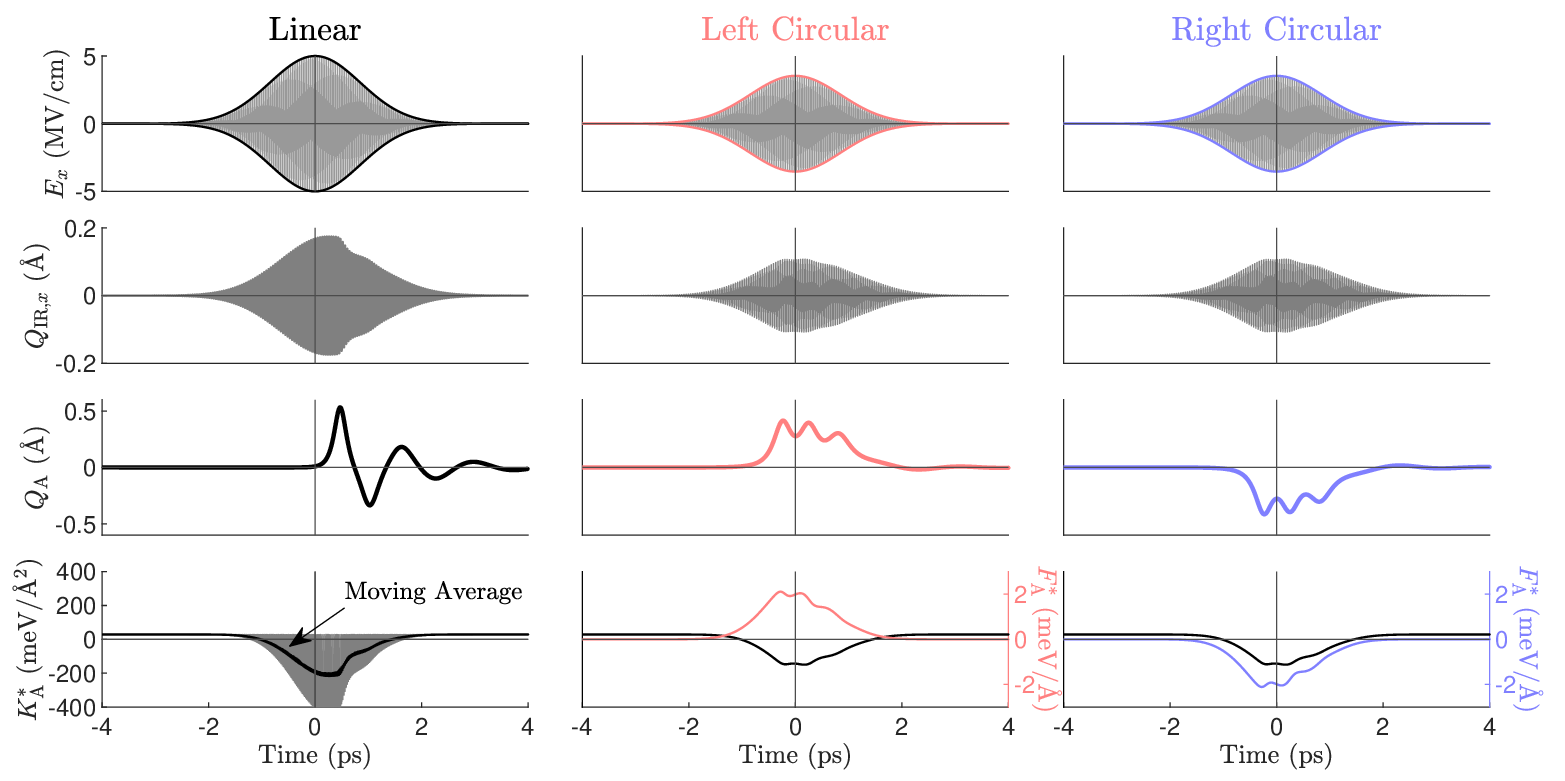}
	\caption{\textbf{Light-induced condensation and directional control of the ferroaxial mode.} Dynamical simulation using light pulse of $E_0$ = 5 MV/cm, $\tau_\text{FWHM}$ = 2 ps, with linear (left column), left-circular (middle column), and right-circular (right column) polarization. We plot the response of IR-active mode (second row) and ferroaxial mode (third row). The bottom row shows the effective force constant $K^*_{\text{A}}$ and unidirectional push $F^*_{\text{A}}$. \label{fig:dynamic}}
\end{figure*}
FIG.~\ref{fig:dynamic} shows a representative dynamical simulation for linear, left- and right-circularly polarized Gaussian pulses with a 5 MV/cm peak electric field $E_0$ and 2 ps pulse duration $\tau_{\text{FWHM}}$. The first column shows the simulated response for light polarized along the $x$-direction (the $y$-direction has an almost identical response). The dynamics of the IR-active phonon show its amplitude nearly saturating, signifying a regime approaching steady-state, a consequence of $\tau_{\text{FWHM}} \approx 1/\gamma_{\text{IR}}$. Near 0.5 ps the IR-active phonon shows an abrupt change in its amplitude. This is associated with a transfer of energy to the ferroaxial phonon $Q_{\text{A}}$. The ferroaxial phonon displaces to large amplitude $\approx$ 0.5 \AA\ through a parametrically amplified process mediated by the biquadratic coupling term $D_{22}$. Notice that although there is a large displacement of the ferroaxial mode, it oscillates about zero until its energy dissipates. This is a result of excitation with linear polarization, in which case there is no preferred direction, so the dynamics of the ferroaxial mode are expected to sample both energy minima \footnote{This statement depends sensitively on the initial condition of the ferroaxial mode and the damping parameter. With appropriate initial conditions and large damping, the ferroaxial mode may prefer a single direction, but this is not likely to be controlled in an experiment with a single linearly polarized pulse.}. In the last row, left column of FIG.~\ref{fig:dynamic}, the time-averaged effective force constant for the ferroaxial mode $K_{\text{A}}^* = K_{\text{A}}+2D_{22}\bm{Q}_{\text{IR}}^2$ becomes negative when the amplitude of the IR-active phonon becomes sufficiently large (recall that $D_{22} < 0$). This suggests a threshold on the strength of the light pulse, only beyond which the ferroaxial phonon becomes unstable and the system launches towards new energy minima. We will elaborate on this threshold phenomenon later in this section.

FIG.~\ref{fig:dynamic} middle and right columns show the simulated response for left- and right-circularly polarized pulses. We show only the $x$-component of the circularly polarized electric field which is $\sqrt{2}$ smaller than the linearly polarized electric field. The $x$-component of the IR-active phonon shows similar behavior for linear and circular polarizations, but the response of the ferroaxial mode is markedly different. For left(right)-circularly polarized light, the ferroaxial mode is pushed to the positive(negative) direction by the NLP, showing that the NLP provides a handle on the direction of the ferroaxial mode. This unidirectional push is exerted by a force $F_{\text{A}}^* = C_c (Q_{\text{IR},x}E_y-Q_{\text{IR},y}E_x)$. This instantaneous, helicity-dependent force $F_{\text{A}}^*$ is plotted together with the instantaneous force constant $K_{\mathrm{A}}^*$ on the last row. We highlight that in contrast to linearly polarized light, the force constant for circularly polarized light does not experience rapid oscillation. Additionally, the ferroaxial mode accesses the first peak at an earlier time (-0.2 ps) than in the linearly polarized case (0.5 ps). This rapid change in the ferroaxial mode amplitude during the rise of the electric field profile is another consequence of the unidirectional push given by NLP.

\begin{figure}[!ht]
    \centering
    \includegraphics[width=0.5\textwidth]{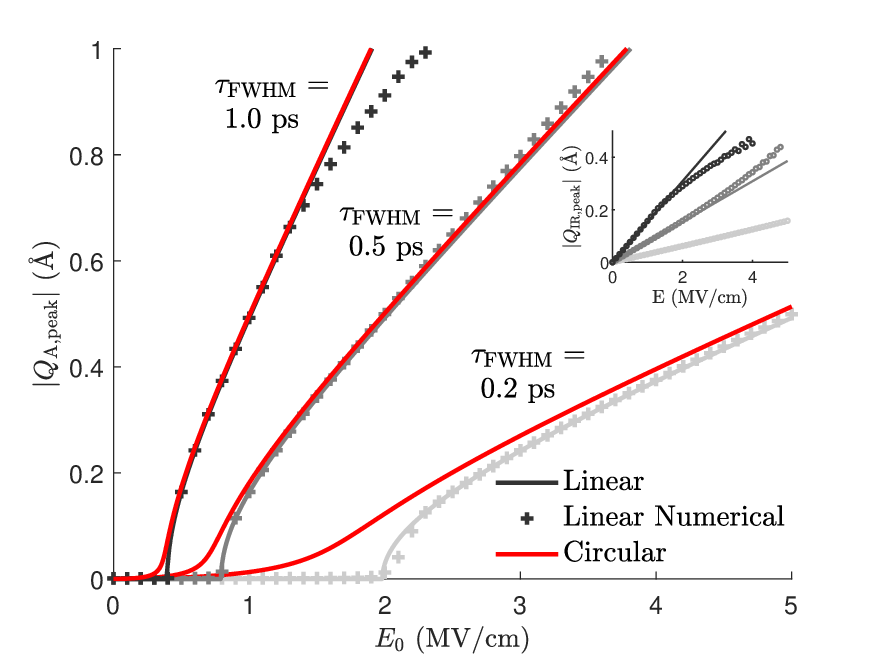}
    \caption{\textbf{Threshold response of the ferroaxial mode.} Peak ferroaxial mode amplitude $\vert Q_{\text{A,peak}}\vert$ vs peak electric field $E_0$. We use Gaussian wave packets with $\tau_{\text{FWHM}}$ = 0.2 ps, 0.5 ps, and 1 ps. The gray(red) curves are the analytical results for linearly(circularly) polarized light. The + symbols represent simulated results using linearly polarized light. Inset:  $\vert \bm{Q}_{\text{IR,peak}}\vert$ vs $E_0$ for linearly polarized light with $\tau_{\text{FWHM}}$ = 0.2 ps, 0.5 ps, and 1 ps from bottom to top. The solid lines and dots are the analytical and simulated results respectively.}
    \label{fig:threshold}
\end{figure}

Next, we investigate the threshold phenomena. In the absence of any excitation, the effective force constant $K_{\text{A}}^* = K_{\text{A}}+2D_{22}\bm{Q}^2_{\text{IR}} = K_{\text{A}}$ is positive, the free energy has one minimum at $Q_{\text{A}}=0$, and the system is in the parent equilibrium phase. When $D_{22}<0$, once the amplitude of the IR-active mode excited by a light pulse crosses a certain threshold, $K_{\text{A}}^*$ becomes negative. $Q_{\text{A}}=0$ is no longer a minimum and the system falls into new minima with finite $Q_{\text{A}}$ (FIG.~\ref{fig:main}d). In the following, we study this threshold phenomenon by solving the minima of the free energy $\mathcal{F}$:
\begin{subequations}
\begin{eqnarray}
\label{eqn:dF}
    \partial\mathcal{F}/\partial \bm{Q}_{\text{IR}}^\top &\approx& \left(K_{\text{IR}}+ 2 D_{22} Q_{\text{A}}^2\right)\bm{Q}_{\text{IR}}  -\tilde{Z}^*\bm{E}(t) \label{eqn:dFdIR} \\ \nonumber &\approx &K_{\text{IR}}\bm{Q}_{\text{IR}}-\tilde{Z}^*\bm{E}(t) = 0 \\
    \partial\mathcal{F}/\partial Q_{\text{A}} & =&\left(K_{\text{A}}+2D_{22} \bm{Q}_{\text{IR}}^2 \right)Q_{\text{A}} + D_{\text{A}}Q_{\text{A}}^3\label{eqn:dFdA}\\ \nonumber &+& C_c(Q_{\text{IR},x}E_y(t)-Q_{\text{IR},y}E_x(t)) = 0 
\end{eqnarray} 
\end{subequations}
In Eqn.~\ref{eqn:dFdIR}, we ignore the effect of $Q_{\text{A}}$ on the IR-active phonon's force constant. This is expected to be a good approximation as long as $K_{\text{IR}} \gg 2 D_{22} Q_{\text{A}}^2$. From Eqn.~\ref{eqn:dFdIR} we find the time-averaged (denoted by $\langle \cdots \rangle$) and peak IR-active mode amplitude when resonantly excited:
\begin{equation}
	\langle  \bm{Q}_{\text{IR}}^2 \rangle = \frac{\tilde{Z}^{*2}\langle \bm{E}^2(t) \rangle}{K_{\text{IR}}^2} = \frac{1}{2}\frac{\tilde{Z}^{*2}\left(2\eta E_0 \omega_{\text{IR}}\tau_{\text{FWHM}}\right)^2}{K_{\text{IR}}^2} 
\end{equation}
and $\vert \bm{Q}_{\text{IR,peak}}\vert = \sqrt{2\langle \bm{Q}_{\text{IR}}^2\rangle}$. Here $E_0^2 = E_{0,x}^2+E_{0,y}^2$ is the peak electric field, and $\eta$ is a factor that characterizes the shape of wave packet, with $\eta=\sqrt{(\pi/2)^3/(2\ln 2)}$ for Gaussian wave packet. 

\begin{figure*}[!ht]
	\centering
	\includegraphics[width=\textwidth]{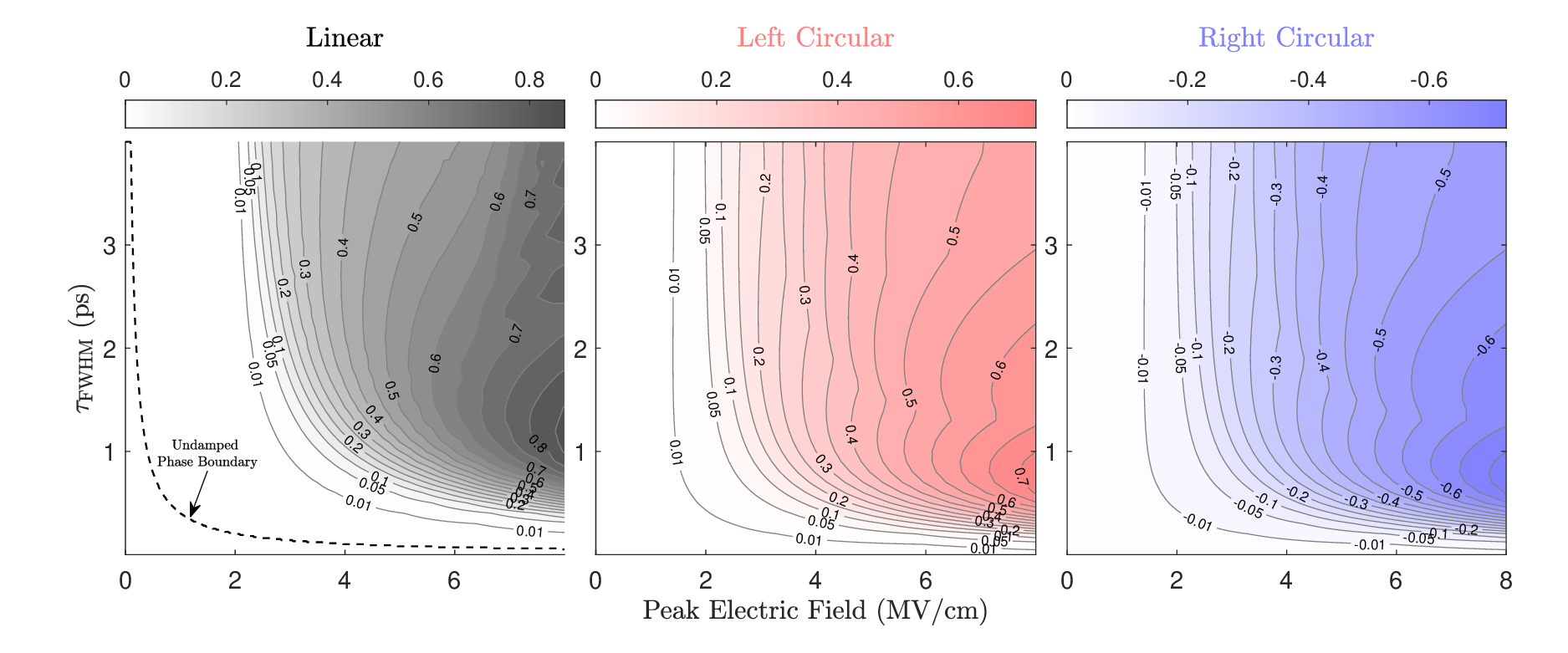}
	\caption{\textbf{Light-induced ferroaxial phase boundary for linearly and circularly polarized light.} Contour plot of the peak ferroaxial mode amplitude $Q_{\text{A,peak}}$ in \AA\ for light pulses of different duration $\tau_{\text{FWHM}}$ and peak electric field $E_0$. We take the absolute value of $Q_{\text{A,peak}}$ for the linear case.}
	\label{fig:phase_diagram}
\end{figure*}

With linearly polarized light, the NLP does not contribute, i.e. $C_c( Q_{\text{IR},x}E_y-Q_{\text{IR},y}E_x) = 0$, and the solution to Eqn. \ref{eqn:dFdA} is simply
\begin{equation} \label{eqn:threshold_QA}
    \langle |Q_{\text{A}}|\rangle = \text{Re} \sqrt{\frac{-\left(K_{\text{A}}+2D_{22}\langle \bm{Q}_{\text{IR}}^2\rangle\right)}{D_{\text{A}}}}
\end{equation}
and $\vert Q_{\text{A,peak}}\vert = \sqrt{2}\langle \vert Q_{\text{A}}\vert\rangle$. Since $\vert Q_{\text{A,peak}}\vert$ is a function of $\langle \bm{Q}_{\text{IR}}^2\rangle$, which depends on both $E_0$ and $\tau_{\text{FWHM}}$, we plot the analytical solutions of $\vert Q_{\text{A,peak}}\vert$ as gray lines in FIG.~\ref{fig:threshold} for pulses with $E_0$ ranging from 0 MV/cm to 5 MV/cm and $\tau_{\text{FWHM}}$ = 0.2 ps, 0.5 ps, and 1.0 ps. The peak $|Q_{\text{A}}|$ obtained from dynamical simulation (shown as +) match the analytical solutions very well, except some deviation at higher $E_0$ for $\tau_{\text{FWHM}}$ = 1.0 ps, where $Q_\text{A}$ is large and our assumption $K_{\text{IR}} \gg 2 D_{22} Q_{\text{A}}^2$ starts to break down. As expected, $\vert Q_{\text{A,peak}}\vert$ becomes nonzero once the light pulse is strong enough to generate sufficiently large IR-active phonon motion. For very short pulses with $\tau_{\text{FWHM}}$ = 0.2 ps, we need a $E_0$ of at least 2 MV/cm. For longer pulses with $\tau_{\text{FWHM}}$ = 0.5 ps or 1 ps, the threshold $E_0$ is reduced to less than 1 MV/cm.

For circularly polarized light resonantly exciting the IR-active phonon, we consider the ideal situation where $Q_{\text{IR},x}$ and $E_y$, and similarly, $Q_{\text{IR},y}$ and $E_x$, are in-phase, and they reach the maximum simultaneously. This leads to the cubic equation:
\begin{equation} \label{eqn:cubic}
	\begin{split}
 D_{\text{A}}\langle Q_{\text{A}}^3 \rangle + \left(K_{\text{A}}+2D_{22} \langle\bm{Q}_{\text{IR}}^2 \rangle\right)\langle Q_{\text{A}}\rangle \\ \pm \frac{1}{2}C_c\vert\bm{Q}_{\text{IR,peak}}\vert E_0 = 0
 \end{split}
\end{equation}
where the sign of the last term depends on the helicity of light. We take the real solution which is in the direction favored by the NLP and plot $ Q_{\text{A,peak}} = \sqrt{2}\langle Q_{\text{A}}\rangle$ as solid red lines in FIG.~\ref{fig:threshold}. Compared to the linearly polarized case, the threshold $E_0$ is lowered and there is no longer a hard cutoff below which $Q_{\text{A}}=0$. This is because the NLP tilts the potential, such that the energy minimum shifts away from $Q_{\text{A}}=0$, effectively creating a unidirectional push to assist the excitation of ferroaxial mode, even for small peak electric fields.

To support the above analysis, we model the threshold phenomenon numerically by extracting the peak ferroaxial mode amplitude $Q_{\text{A,peak}}$ from dynamical simulations with damping as in FIG.~\ref{fig:dynamic}, with the same model parameters. We simulate the response of the ferroaxial mode for linearly and circularly polarized light pulses of frequency 23.7 THz, as before, with peak electric field $E_0$ ranging from 0 MV/cm to 8 MV/cm and pulse duration $\tau_{\text{FWHM}}$ ranging from 0 ps to 4 ps, plotting $Q_{\text{A,peak}}$ as contours in FIG.~\ref{fig:phase_diagram}. 

For linear polarization, we find the peak ferroaxial amplitude to be sensitive to the initial conditions because the local maximum at $Q_{\text{A}}=0$ is an unstable equilibrium point in the lattice potential, see SI IV. To mitigate this issue for sensible comparison with experimental work, we average over the phase space of initial conditions with small displacements and small velocities. In FIG.~\ref{fig:phase_diagram} left, the numerical phase boundary (solid lines) is shifted from the analytic phase boundary obtained from solving Eqn. \ref{eqn:threshold_QA} (dashed line). This is due primarily to the damping of the IR-active phonon, see SI III. 

Threshold phenomena are seen for both linearly and circularly polarized light, i.e. $Q_{\text{A}}$ only becomes sizeable when the light pulse is strong enough. Consistent with the analytic results in the absence of IR-active phonon damping in FIG. \ref{fig:threshold}, the phase boundary is most sensitive to the light's polarization when the pulse duration is short. That is, a lower peak electric field is needed with circularly polarized light, with the effect most noticeable at short durations ($<$ 1 ps). 

Another common feature is that for a high peak electric field, the maximum peak ferroaxial amplitude occurs at a relatively short pulse duration (1.2 ps for linear and 0.8 ps for circular polarization). This is attributed to the coupling between IR-active and ferroaxial phonons. For longer pulse durations, the effective force constant $K_{\text{IR}}^*=K_{\text{IR}}+2D_{22}Q_{\text{A}}^2$ can deviate substantially from the equilibrium value as $Q_{\text{A}}$ grows, lowering the frequency of the IR-active phonon. In experiment, this may be partly managed with a chirped pulse that is tuned to the response of the IR-active phonon, though we expect this to be impractical since detailed knowledge of the dynamics of $Q_{\text{A}}$ and $\bm{Q}_{\text{IR}}$ will be needed to define the chirp.

\subsection{Switching of ferroaxial domains \label{sec:switch}}
\begin{figure*}[!ht]
	\centering
	\includegraphics[width=\textwidth]{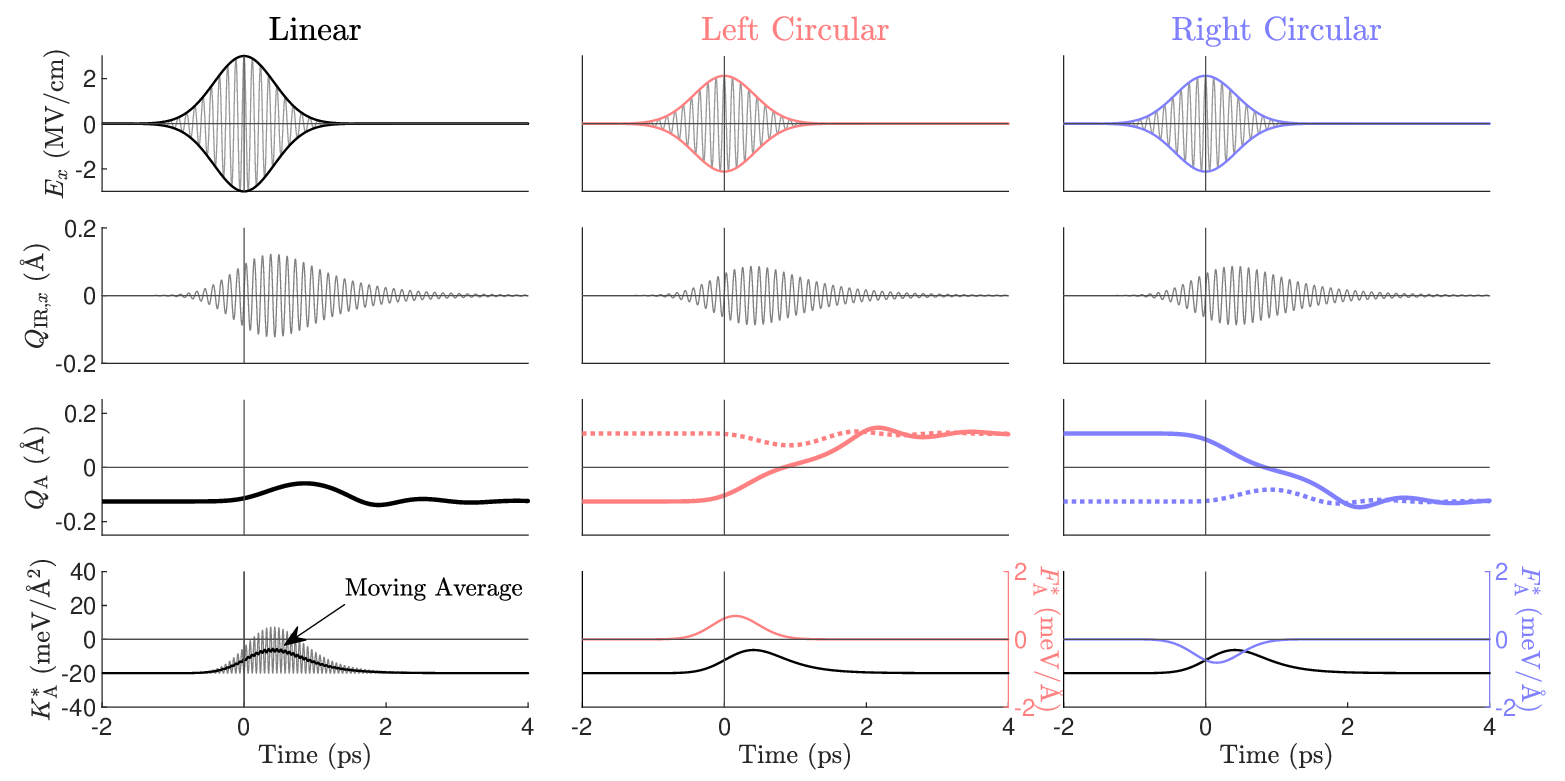}
	\caption{\textbf{Ferroaxial domain switching with circularly polarized light.} Dynamical simulation using $E_0$ = 3 MV/cm and $\tau_{\text{FWHM}}$ = 1 ps resonant with the 8.6 THz IR-active phonon, for linear (left column), left-circular (middle column) and right-circular (right column) polarizations. The response of the IR-active mode (second row), ferroaxial mode (third row), and its effective force constant and unidirectional push (fourth row) are shown. Both ferroaxial domains converge to a single domain after the pulse, as indicated by the solid and dashed line in the third row.}
	\label{fig:switch}
\end{figure*}
Since the circularly polarized light generates a helicity-dependent unidirectional push to the ferroaxial mode, we explore the possibility of switching the ferroaxial domain when the crystal is already in the ferroaxial phase, e.g. below the ferroaxial phase transition temperature. 

To illustrate this, without loss of generality, we assume the crystal is initially in one of the energy minima, as illustrated in FIG.~\ref{fig:main}e, and there is a finite barrier between the energy minima separating the two ferroaxial domains. We estimate the condition for switching as the point where the barrier height equals zero, noting that thermal and quantum fluctuations make this an approximate upper bound for the switching.

We deduce that the barrier height can be reduced to zero only if
\begin{equation}
	\Delta \equiv -4p^3-27q^2 < 0 \label{eqn:Delta}
\end{equation}
where
\[p = \frac{K_\text{A}+2D_{22}\langle\bm{Q}_{\text{IR}}^2\rangle}{D_{\text{A}}}\quad \text{and}\quad q=\frac{C_c\vert \bm{Q}_{\text{IR,peak}}\vert E_0}{2D_{\text{A}}}\]
depend on the intrinsic mode parameters $D_{22}$ and $C_c$. Thus not all doubly degenerate IR-active modes are able to switch the domain. Circularly polarized light, 1) tilts the double well through the NLP and 2) deepens or shallows the double well depending on the sign of biquadratic coupling coefficient $D_{22}$. For the 23.7 THz mode, we find that the barrier cannot be lowered to zero as the bias generated by NLP is not enough to overcome the deepening of the double well due to its large negative $D_{22}$. We discuss the parameter dependence of light-induced ferroaxial switching in SI V, where the parameters may be sensitive to temperature, an effect not modeled directly in this work. In what follows, we focus on another mode with positive $D_{22}$.

We select the 8.6 THz mode that has slightly positive $D_{22}$ with sizable $C_c$ to demonstrate the domain switching in RbFe(MoO$_4$)$_2$. We choose a small negative value of $K_\text{A}$ = -0.02 eV/\AA$^2$ so that the double well is shallow initially, simulating a scenario where, for example, the system's temperature is just below its critical temperature. In FIG.~\ref{fig:switch}, we compare the response to linearly, left-, and right-circularly polarized light. The linear case does not exhibit switching behavior because the effective force constant is negative throughout the process, so a finite barrier persists. We note that this suggests that the para-axial phase boundary has not been crossed. In the circularly polarized cases, one domain remains unaffected, while the opposite domain switches under the assistance of the NLP which tilts the double well. Our simulations demonstrate the ability to deterministically switch domain and pole the crystal, i.e. from a multi-domain state to a single-domain state, with the direction of the desired single domain chosen by the helicity of light pulse. This scenario can be distinguished from local heating by its coherent, ultrafast structural response, on a picosecond timescale.

\section{Discussion \label{sec:discussion}}
We have shown that the excitation of an IR-active phonon can selectively activate the ferroaxial order parameter through the intrinsic material's response mediated by the anharmonic lattice potential and nonlinear lattice polarizability. How can this IR-light-induced ferroaxial order be detected? The most direct measurement of the induced ferroaxial order can be found via a mid-infrared pump/x-ray probe experiment. In such an experiment, the temporally resolved changes to the structure factor can be measured directly. Our calculations of structure factor changes with respect to the ferroaxial mode show that there are multiple easily detectable Bragg conditions for the focus of such an experiment. More information is in SI VI. For example, the $\left\{11\bar{2}1\right\}$ Bragg peak, a total of twelve symmetry-related points, are degenerate in the parent structure with a structure factor of $\approx$ 20. When the ferroaxial mode is excited, this Bragg peak splits into two six-fold degenerate Bragg peaks whose intensities increase/decrease by about $\left(16\ \right.$\AA$^{-1}  \left.\right)\times Q_\text{A}$, which means that a 0.5 \AA\  ferroaxial mode changes the structure factor by $\approx\pm8$.

Optical probing and even direct imaging of ferroaxial domains have been achieved by multiple authors via second harmonic generation through the electric quadrupole moment \cite{Jin_2019,Yokota_2022,Guo_2023}, which was also recently demonstrated on ultrafast timescales \cite{Luo_2021}. An alternative optical approach was demonstrated in experiments by Hayashida et. al., who spatially resolved the ferroaxial domain via the linear electrogyration effect \cite{Hayashida_2020,Hayashida_2021}. The induced ferroaxial response may also be identified by extending these optical probes of ferroaxial order to mid-infrared pump/optical probe experiments. Additionally, the dynamics of the ferroaxial mode suggest an optical strategy for characterizing the temporal response of $Q_\text{A}$. The time-derivative of $Q_\text{A}$ represents a circular bound current in the material, which couples to the electric field as
\begin{equation}
    \Delta \mathcal{F} = \tilde{\chi} \left[E^*(\omega_1)\times E(\omega_2) \right]\cdot \left[-i \omega_3 Q_\text{A}(\omega_3)\right]
\end{equation}
where the $-i\omega_3$ follows from the time-derivative. This represents an anti-symmetric deviation of the dielectric susceptibility at the frequency of the ferroaxial mode taking the form $\delta\epsilon_{ij} = -i\varepsilon_{ijk}\omega_3\tilde{\chi}Q_{A,k}(\omega_3)$ which may be probed by light at any frequency, analogous to the Faraday effect or magnetic circular dichroism.

In our development of the light-induced ferroaxial order, we have so far ignored the effect of strain. Of course, as the optical energy is deposited into the IR-active phonon, a strain response is also expected in the crystal \cite{Hortensius_2020}. We find that the phase boundary including strain can be modeled with the same analytic form of Eqn.~\ref{eqn:threshold_QA} with renormalized parameters and, therefore has no qualitative changes. The analysis is included in SI II. Although more optical energy is required to induce the ferroaxial phase when strain is considered, the pulse duration and peak field required are within current experimental capabilities \cite{Mankowsky_2017}. 

It is evident that larger magnitude $C_c$ is more favorable for selecting the ferroaxial domains. The necessity of this intrinsic parameter begs the question: Is our approach applicable to a wide range of potential ferroaxial materials or a special feature of RbFe(MoO$_4$)$_2$? Numerical exploration of this problem shows that reducing $C_c$ by an order of magnitude still allows for a single light-induced ferroaxial domain in the early timescale dynamics. For ferroaxial domain switching, decreasing $C_c$ increases the threshold electric field needed for switching. Under such circumstances, nonetheless, strong domain control may still be achieved with a sufficiently intense light pulse for systems very close to the phase boundary, see details in SI IV.

\section{Summary and Outlook \label{sec:summary}}
In summary, we establish through symmetry analysis, a coupling field to ferroaxial order naturally expressed through the nonlinear lattice polarizability. This provides a novel optical strategy to control ferroaxial order, a feature not accessible through other conventional fields. Transiently induced single-domain ferroaxial order above the ferroaxial phase transition temperature and switching/poling of ferroaxial domains below the phase transition temperature are both enabled by helicity control of mid- and far-IR light. Our simulations on RbFe(MoO$_4$)$_2$ show the feasibility of our strategy for controlling ferroaxial order on ultrafast timescale. With this finding, we anticipate that continued study of light-matter interactions in the mid-(far-)infrared will give new insights into the control of novel orders in condensed phases of matter. 

\section*{ACKNOWLEDGMENTS \label{sec:acknowledgements}}
We thank Nicole Benedek, Craig Fennie, Jiaoyang Zheng, and Ankit Disa for fruitful discussions. G.K. acknowledges support from the Cornell Center for Materials Research with funding from the NSF MRSEC program (Grant No. DMR-1719875). Computational resources are provided by the Cornell Center for Advanced Computing. 

\appendix
\section{Computational Method\label{app:dft}}
We compute structural properties and extract model parameters using density functional theory with a Hubbard U correction (DFT+U). The DFT+U calculation is implemented in the Vienna Ab-initio Simulation Package (VASP) \cite{vasp_1,vasp_2,vasp_3} with PBEsol exchange-correlation functionals \cite{PBEsol} and projector augmented wave pseudopotentials \cite{vasp_paw}. We find structural convergence with a kinetic energy cutoff of 550 eV and a k-mesh of 6$\times$6$\times$4 for the primitive cell. We find no qualitative change in the structural and magnetic properties for reasonable variation of the Hubbard U and J parameters. We therefore use U = 4 eV, and J = 0.9 eV, in accordance with a previous study \cite{Cao_2014}. In our calculations, we use the ferromagnetically ordered phase to calculate the structure and model parameters, as the symmetry is more representative of the crystal above the magnetic phase transition temperature where the ferroaxial phase transition is seen. Additional tests have shown that the qualitative results of our study are insensitive to the choice of magnetic order. Structures are converged to within 1$\times$10$^{-3}$ eV/\AA\ force threshold on each atom. We note that in PBEsol, the high-temperature parent structure (P$\bar{3}$m1) is the ground state. This is due to a strong coupling between the in-plane lattice constant and the ferroaxial mode and a small over-estimation of the in-plane lattice constant in PBEsol. The structures are visualized using VESTA \cite{Momma:db5098}. More details can be found in SI I.

\section{Condition for nonzero NLP push to ferroaxial modes \label{app:NLP}} 

\begin{figure*}[!ht]
    \subfloat[No $Q_{\text{A}}$]{\includegraphics[width=0.49\textwidth]{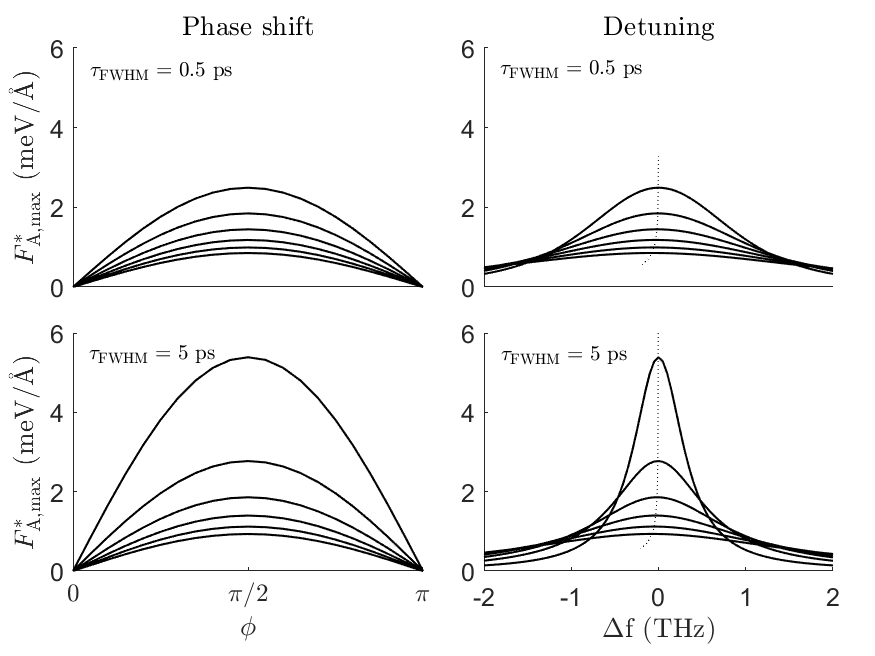}} \hfill
    \subfloat[With coupling to $Q_{\text{A}}$]{\includegraphics[width=0.49\textwidth]{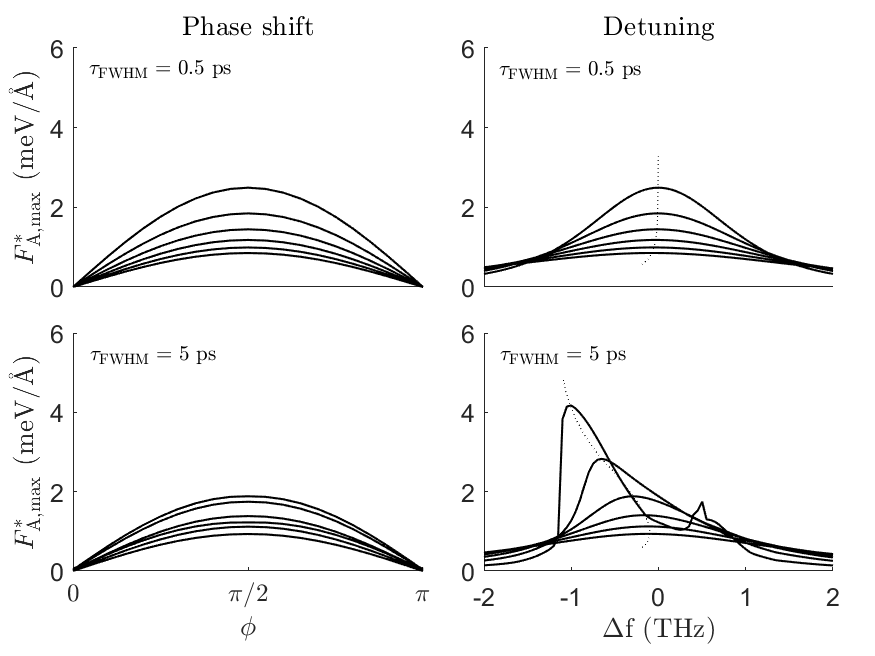}}
    \caption{$F^*_{\text{A,max}}$ with varying $\phi$ and light frequency detuning, for $\tau_{\text{FWHM}}$ = 0.5 ps or 5 ps. The 6 curves in each panel correspond to $\gamma_{\text{IR}}$ = 2 THz, 4 THz, 6 THz, 8 THz, 10 THz, and 12 THz. The dotted lines trace the location of the peak.}
    \label{fig:AA}
\end{figure*}

We show that the NLP term that drives the ferroaxial mode, which is present in all point groups leading to ferroaxial phases, can be accessed in the plane wave (long pulse duration) and impulsive limits (short pulse duration), provided there is some ellipticity in the light polarization and the excitation is near-resonance.

\subsection{Long pulse duration -- Plane wave limit}
The displacement of IR-active mode is described by the equation 
\[\ddot{Q}_{\text{IR},\{x,y\}}+2\gamma_{\text{IR}}\dot{Q}_{\text{IR},\{x,y\}}+\omega_0^2Q_{\text{IR},\{x,y\}} = -\frac{\tilde{Z}^*E_{\{x,y\}}}{M_\text{IR}}\]
\noindent
where the displacement is driven by a monochromatic light propagating along $\hat{z}$ with the general form
\[\left\{
\begin{array}{l}
	E_x = E_{0x}\cos{\omega t}\\
	E_y = E_{0y}\cos{(\omega t + \phi)}.
\end{array} \right.\]
Solving the differential equation of a damped driven oscillator and omitting the exponentially decaying (transient) solution, we obtain
\[\left\{
\begin{array}{l}
	Q_{\text{IR},x} = -CE_{0x}\cos{(\omega t + \delta)}\\
	Q_{\text{IR},y} = -CE_{0y}\cos{(\omega t + \phi + \delta)}
\end{array} \right.\]
where \[C = \frac{\tilde{Z}^*}{M_\text{IR}\sqrt{\left[(\omega_0^2-\omega^2)^2+4\gamma_{\text{IR}}^2\omega^2\right]}}\ \textrm{and}\ \tan\delta = \frac{4\gamma_{\text{IR}}\omega}{\omega^2-\omega_0^2}.\]

\noindent
C is maximized when
\[\omega = \sqrt{\omega_0^2-2\gamma_{\text{IR}}^2}\]
Inserting this result into the expression $Q_{\text{IR},x}E_y-Q_{\text{IR},y}E_x$, we have
\begin{align*}
    & Q_{\text{IR},x}E_y-Q_{\text{IR},y}E_x \\
    = & \frac{1}{2}CE_{0,x}E_{0,y}\left[\cos{(\delta-\phi)} - \cos{(\delta+\phi)} \right]\\
    = & CE_{0,x}E_{0,y}\sin{\delta}\sin{\phi}
\end{align*}
We can immediately see that the direction of the force on the ferroaxial mode can be controlled by the helicity through $\phi$. 

This expression vanishes when $\delta\rightarrow 0$ (far from resonance) or $\phi=0$ (linearly polarized light). We can see that the expression becomes maximal under two conditions: \text{(1)} If damping $\gamma_{\text{IR}}$ is negligible, $\omega=\omega_0$ so that C is maximized and $\sin\delta = \sin(-\pi/2) = 1$. 
If some damping is present, a slightly lower $\omega$ will maximize the response, which we do not explore here.
(2) $\phi = \pm \pi/2$. When  $E_{0,x}=E_{0,y}$, this condition corresponds to the left/right circularly polarized light. 

\subsection{Short pulse duration -- Impulsive limit}
When we excite the mode with an extremely short pulse, the damping term can be ignored and the equations read 
\[
    \Ddot{Q}_{\text{IR},\{x,y\}}+\omega_0^2Q_{\text{IR},\{x,y\}} = -\frac{\tilde{Z}^*E_{\{x,y\}}}{m_\text{IR}}
\]
with the electric field taking the form
\[
    \left\{
\begin{array}{l}
	E_x = E_{0x}\Phi(t)\cos{\omega t}\\
	E_y = E_{0y}\Phi(t)\cos{(\omega t + \phi)}
\end{array} \right.
\]
where $\Phi(t)$ is the envelope function that describes the shape of the pulse (for example, a Gaussian $\Phi(t)=e^{-t^2/2\sigma^2}$). Solving for the components of the IR-active phonon, we find,
\begin{align*}Q_{\text{IR},x}(t) = & \frac{E_{0x}}{\omega_0}\int_{-\infty}^t d\tau \Phi(\tau)\cos{(\omega\tau)}\sin{[\omega_0(t-\tau)]}\\
= & \frac{E_{0x}}{\omega_0}\sin{(\omega_0t)}\int_{-\infty}^t d\tau \Phi(\tau)\cos{(\omega_0\tau)}\cos{(\omega\tau)}\\
-& \frac{E_{0x}}{\omega_0}\cos{(\omega_0t)}\int_{-\infty}^t d\tau \Phi(\tau)\sin{(\omega_0\tau)}\cos{(\omega\tau)}
\end{align*}
\begin{align*}Q_{\text{IR},y}(t) = & \frac{E_{0y}}{\omega_0}\int_{-\infty}^t d\tau \Phi(\tau)\cos{(\omega\tau+\phi)}\sin{[\omega_0(t-\tau)]}\\
= & \frac{E_{0y}}{\omega_0}\sin{(\omega_0t)}\int_{-\infty}^t d\tau \Phi(\tau)\cos{(\omega_0\tau)}\cos{(\omega\tau+\phi)}\\
-& \frac{E_{0y}}{\omega_0}\cos{(\omega_0t)}\int_{-\infty}^t d\tau \Phi(\tau)\sin{(\omega_0\tau)}\cos{(\omega\tau+\phi)}
\end{align*}

We now consider the resonant case $\omega=\omega_0$. Defining $C_1\equiv 1/\omega_0\int_{-\infty}^t d\tau \Phi(\tau)\cos^2{(\omega_0\tau)} > 0$, $C_2\equiv 1/\omega_0\int_{-\infty}^t d\tau \Phi(\tau) \sin^2{(\omega_0\tau)} > 0$ and $C'\equiv1/\omega_0\int_{-\infty}^t d\tau \Phi(\tau)\cos{(\omega_0\tau)}\sin{(\omega_0\tau)}$, we can rewrite $Q_{\text{IR},x}$ and $Q_{\text{IR},y}$ as
\[Q_{\text{IR},x}(t)=E_{0,x}\left[C_1\sin{(\omega_0 t)}-C'\cos{(\omega_0 t)}\right]\]
\begin{align*}Q_{\text{IR},y}(t) 
= E_{0,y}\sin{(\omega_0 t)}\left[C_1\cos{\phi}-C'\sin{\phi}\right] \\+ E_{0,y}\cos{(\omega_0 t)}\left[C_2\sin{\phi}-C'\cos{\phi}\right]
\end{align*}
As a result
\begin{align*}
    &Q_{\text{IR},x}E_y-Q_{\text{IR},y}E_x\\
    = &-E_{0,x}E_{0,y}\Phi(t)\sin{\phi} \\ \times&\left[C_1\sin^2{(\omega_0t)}+C_2\cos^2{(\omega_0t)}-2C'\cos{\omega_0t}\sin{\omega_0t}\right]\\
\end{align*}
is proportional to $\sin{\phi}$, reaching a maximum when $\phi=\pm\pi/2$, i.e. circularly polarized light if $E_{0,x}=E_{0,y}$.

\subsection{Long and Short Duration Pulses Limits -- Numerical Verification}
To better understand the lattice mode behavior under different light pulses and to verify our reasoning above, we perform dynamical simulation using pulses of varying frequencies and phase shifts between components of the electric field $E_x$ and $E_y$. To illustrate the primary results, we use a pulse with a peak electric field of 5 MV/cm with E$_{0,x}$ = E$_{0,y}$ and $\tau_{\text{FWHM}}$ = 0.5 ps (impulsive limit) or 5 ps (plane-wave limit). We vary $\gamma_{\text{IR}}$ to study the effect of damping on the unidirectional push to the ferroaxial mode via the NLP. We extract the maximum unidirectional push in the simulation $F^*_{\text{A,max}}$=Max$_t\left\{C_c(Q_{\text{IR},x}(t)E_y(t)-Q_{\text{IR},y}(t)E_x(t))\right\}$ and plot in FIG.~\ref{fig:AA}.

In both the impulsive and plane wave limits, $F^*_\text{A,max}$ peaks when $\phi=\pi/2$. For the short pulse, $F^*_\text{A,max}$ also peaks when the driving frequency equals the IR frequency. For the longer pulse, the peak appears at a slightly lower frequency, due primarily to the back action of the ferroaxial mode on the IR-active mode through $D_{22}$ -- decreasing the instantaneous IR-active phonon frequency. 

\pagebreak
\bibliography{ref}
\end{document}


\title{SUPPLEMENTAL INFORMATION \\
Optical Control of Ferroaxial Order}
\date{\today}
\author{Zhiren He}
\email{zhiren.he@unt.edu}
\affiliation{School of Applied and Engineering Physics, \\ Cornell University, Ithaca, New York 14853, USA}
\affiliation{Department of Physics, University of North Texas, Denton, TX 76203, USA}
\author{Guru Khalsa}
\email{guru.khalsa@unt.edu}
\affiliation{Department of Materials Science and Engineering, \\ Cornell University, Ithaca, New York 14853, USA}
\affiliation{Department of Physics, University of North Texas, Denton, TX 76203, USA}
\maketitle

\section{Computational Details \label{supp:computation}} 
\subsection{Structure}
We present the relaxed lattice and atomic positions in DFT+U in TABLE~\ref{tab:struct}, along with the structural parameters determined by single crystal X-ray diffraction at 295 K and 110 K \cite{Wakowska_2010}. Both calculated lattice constant and atomic positions in the parent P$\bar{3}$m1 phase are very close to the experimental value at 295 K. The structure at 110 K is in the lower symmetry P$\bar{3}$ phase. Compared with 295 K structure, the \textit{a} and \textit{c} lattice constant shrink by 1.1 \% and 0.5\%, respectively, and the 3 planar oxygen atoms in the MoO$_4$ tetrahedron shift by approximately 0.27 \AA\ while other atoms displace little. 
\begingroup \squeezetable
\begin{table}[!hb]
    \centering
    \begin{ruledtabular}
    \begin{tabular}{ccccccccccc}
        & Space group & a (\AA) & c (\AA) & V (\AA$^3$) & Site & Rb & Fe & 2Mo & 2O$_a$ & 6O$_p$  \\ \cmidrule(lr){1-2} \cmidrule(lr){3-5} \cmidrule(lr){6-11}
        \multirow{3}{*}{DFT} & \multirow{3}{*}{P$\bar{3}$m1} & \multirow{3}{*}{5.691} & \multirow{3}{*}{7.440} & \multirow{3}{*}{208.7} & x & 0 & 0 & 1/3 & 1/3 & 0.1615  \\ 
        ~ & ~ & ~ & ~ & ~ & y & 0 & 0 & 2/3 & 2/3 & -0.1615  \\ 
        ~ & ~ & ~ & ~ & ~ & z & 0.5 & 0 & 0.2319 & 0.464 & 0.1580  \\ \cmidrule(lr){1-2} \cmidrule(lr){3-5} \cmidrule(lr){6-11}
        \multirow{3}{*}{295 K} &\multirow{3}{*}{P$\bar{3}$m1} & \multirow{3}{*}{5.669} & \multirow{3}{*}{7.492} & \multirow{3}{*}{208.6} & x & 0 & 0 & 1/3 & 1/3 & 0.1620  \\ 
        ~ & ~ & ~ & ~ & ~ & y & 0 & 0 & 2/3 & 2/3 & -0.1620  \\ 
        ~ & ~ & ~ & ~ & ~ & z & 0.5 & 0 & 0.2293 & 0.458 & 0.1577  \\ \cmidrule(lr){1-2} \cmidrule(lr){3-5} \cmidrule(lr){6-11}
        \multirow{3}{*}{110 K} & \multirow{3}{*}{P$\bar{3}$} & \multirow{3}{*}{5.607} & \multirow{3}{*}{7.452} & \multirow{3}{*}{202.9} & x & 0 & 0 & 1/3 & 1/3 & 0.1127  \\
        ~ & ~ & ~ & ~ & ~ & y & 0 & 0 & 2/3 & 2/3 & -0.1617  \\
        ~ & ~ & ~ & ~ & ~ & z & 0.5 & 0 & 0.2309 & 0.4614 & 0.1585  \\ 
    \end{tabular}
    \end{ruledtabular}
    \caption{Relaxed lattice and atomic position using DFT, compared with experiments at 295 K and 110 K. O$_a$ and O$_p$ are the apical and planar oxygens in MoO$_4$ tetrahedron. }
    \label{tab:struct}
\end{table}
\endgroup
\subsection{DFT+U and magnetism}
We test the effect of the U-parameter in our DFT+U calculation, listed in rows 2-4 of TABLE~\ref{tab:U}. Variation of U from 2 eV to 6 eV causes only minor changes in lattice parameters and phonon frequencies. In rows 5-7 of TABLE~\ref{tab:U} we check if magnetism alters the structural properties. The experimental N\'eel temperature is 3.8 K \cite{TN}, much lower than the ferroaxial transition temperature of 195 K, therefore, we expect the effect of magnetism to be negligible. This is confirmed by comparing other magnetic configurations like $\uparrow\downarrow$ (AFM in-plane, cell size = 2), $\uparrow\downarrow\downarrow$ (1 up, 2 down in-plane, cell size = 3) and $\uparrow\downarrow$ OOP (AFM out-of-plane, cell size = 2) with $\uparrow$ (FM). 
\begingroup \squeezetable
\begin{table}[!htb]
    \centering
    \begin{ruledtabular}
    \begin{tabular}{cccccccc}
         \makecell{Magnetic \\Configuration} & U (eV) & $a$ (\AA) & $c$ (\AA)  & \multicolumn{4}{c}{Low Freq Phonons at $\Gamma$ (THz)} \\ \midrule
         \multirow{3}{*}{$\uparrow$} & 2 & 5.695 & 7.432  & 0.360 & 1.616 & 1.647 & 3.041 \\
          & 4 & 5.691 & 7.440  & 0.278 & 1.633 & 1.676 & 3.101 \\
          & 6 & 5.684 & 7.446 & 0.470 & 1.600 & 1.673 & 3.077 \\ \midrule
          $\uparrow\downarrow$ & 4 & $\approx$ 5.69 & 7.438 &  0.408 & 1.600 & 1.660 & 3.062 \\
          $\uparrow\downarrow\downarrow$ & 4 & 5.691 & 7.439 &  0.473 & 1.598 & 1.663 & 3.060 \\
          $\uparrow\downarrow$ OOP & 4 &  5.691 & 7.440 &  0.407 & 1.600 & 1.660 & 3.062 
    \end{tabular}
    \caption{Lattice parameters and 4 lowest-frequency phonons at $\Gamma$-point calculated with different U and magnetic configurations. J is fixed to 0.9 eV. No spin-orbit-coupling is included.\label{tab:U}}
    \end{ruledtabular}
\end{table}
\endgroup
\subsection{Fitting model parameters}
\begin{figure}[!htb]
    \subfloat[]{\includegraphics[width=0.49\textwidth]{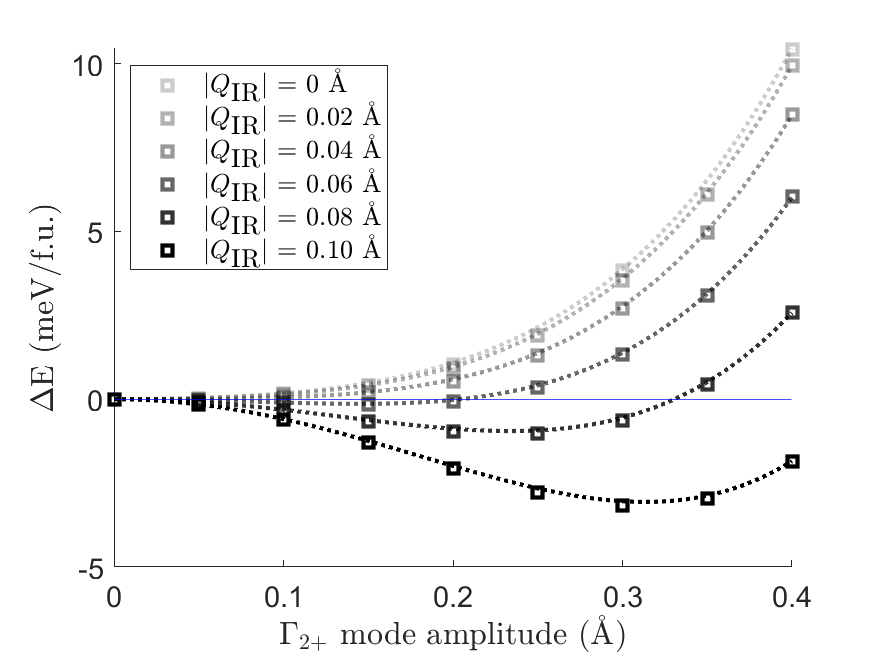}} \hfill
    \subfloat[]{\includegraphics[width=0.49\textwidth]{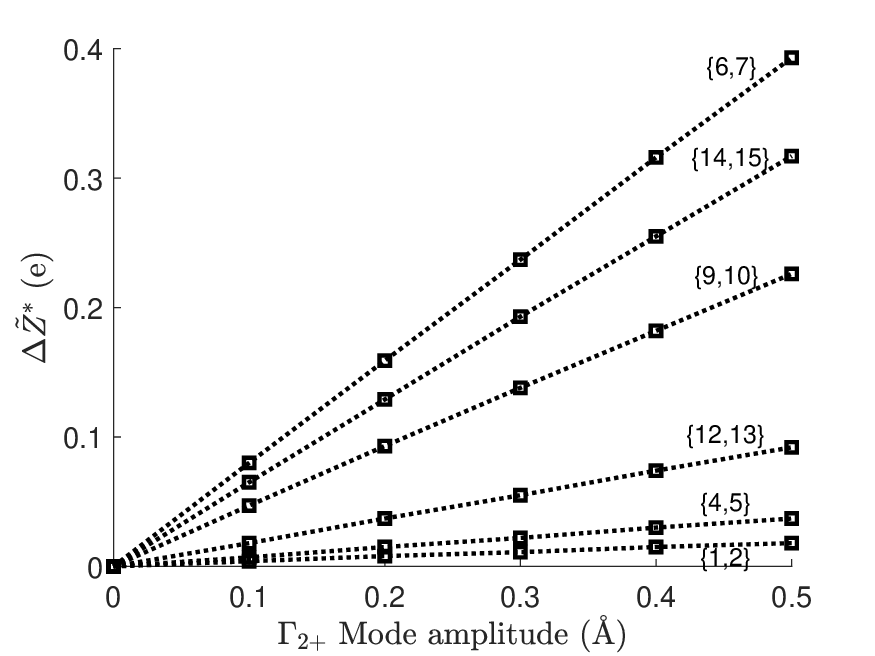}}
    \caption{(a) Fit to find $K_{\text{A}}$, $D_{\text{A}}$ and $D_{22}$. The IR-active mode shown here is IR No. 9. Squares are the DFT results and dotted curves are the fittings. (b) Fit to find the nonlinear polarizability $C_c$.}
    \label{fig:fit}
\end{figure}

Here we present additional details on the extracted model coefficients through fitting DFT computed results. The free energy has the following terms
\begin{eqnarray*}
    \mathcal{F} =& \frac{1}{2}K_{\text{IR}} \bm{Q}_{\text{IR}}^2 \\
    +& \frac{1}{2}K_{\text{A}}Q_{\text{A}}^2 
    + \frac{1}{4} D_{\text{A}}Q_{\text{A}}^4\\
    +& D_{22} \bm{Q}_{\text{IR}}^2Q_{\text{A}}^2\\ -&
    \tilde{Z}^* \left( Q_{\text{IR},x} E_x + Q_{\text{IR},y} E_y \right) \\
    -& C_c Q_{A}\left(Q_{\text{IR},x} E_y - Q_{\text{IR},y} E_x \right)
\end{eqnarray*}
$K_{\text{IR}}$ can be computed directly from the IR-active phonon frequency and the mode-effective charges $\Tilde{Z}^*$ are deduced from the Born effective charges, shown in TABLE~\ref{tab:IR}. The coefficients $K_{\text{A}}$, $D_{\text{A}}$, and $D_{22}$ may be found by fitting a series of DFT static energy calculations. In each step, we freeze in some amplitude of IR-active modes (0 \AA -- 0.1 \AA) to the fully relaxed parent structure and compute the energy with increasing amplitude of the ferroaxial mode (0 \AA -- 0.4 \AA). The change in energy due to the added ferroaxial mode is  
$\Delta E =  \frac{1}{2}K_{\text{A}}Q_{\text{A}}^2 
+ \frac{1}{4} D_{\text{A}}Q_{\text{A}}^4
+ D_{22} \bm{Q}_{\text{IR}}^2Q_{\text{A}}$
(plotted as squares in FIG.~\ref{fig:fit}a). The coefficients are then extracted from the fitting. 

Lastly, we determine the value of $C_c$. From the expression of the NLP, 
\[C_cQ_\text{A} = \frac{\partial\Delta P_{\text{NL},y}}{\partial Q_{\text{IR},x}} = \Delta\tilde{Z}^*_{xy}= -\frac{\partial\Delta P_{\text{NL},x}}{\partial Q_{\text{IR},x=y}} = -\Delta\tilde{Z}^*_{yx}\]
We add the ferroaxial mode to the fully relaxed parent structure and extract the IR-active mode-effective charge $\Tilde{Z}^*$ from the DFT output. In FIG.~\ref{fig:fit}b, we plot the absolute change of $\Tilde{Z}^*$ in the $x(y)$ direction for IR-active mode in the $y(x)$ direction parameterized by ferroaxial mode amplitude. The slope of the line gives $C_c$.

\section{Lattice effect \label{supp:strain}}
In this section, we quantify the effect of the lattice by including elastic energy, and coupling of strain ($\epsilon$) with the ferroaxial and IR-active phonons in the free energy. Up to third order, considering both in-plane ($\epsilon_a = \epsilon_{11}$) and out-of-plane ($\epsilon_c = \epsilon_{33}$) strains, these can be written as
\[
\begin{split}
    F =& \frac{1}{2}K_{\text{IR}}\bm{Q}_{\text{IR}}^2 
    + \frac{1}{2}K_{\text{A}}Q_{\text{A}}^2 +\frac{1}{4} D_{\text{A}}Q_{\text{A}}^4 
    + D_{22}\bm{Q}_{\text{IR}}^2Q_{\text{F}}^2
    - \Delta \bm{P} \cdot\bm{E} \\
    + & \frac{1}{2}K_{a}\epsilon_a^2 + \frac{1}{2}K_{c}\epsilon_c^2 \quad \text{(Elastic energy)}\\
    + & C_{a\text{A}}\epsilon_aQ_{\text{A}}^2 + C_{c\text{A}}\epsilon_cQ_{\text{A}}^2 \quad \text{(Strain-ferroaxial coupling)}\\
    + & C_{a\text{IR}}\epsilon_a\bm{Q}_{\text{IR}}^2 + C_{c\text{IR}}\epsilon_c\bm{Q}_{\text{IR}}^2 \quad \text{(Strain-IR coupling)}
\end{split}
\]
where the coefficients $K_{a}$, $K_{c}$, $C_{a\text{A}}$, $C_{c\text{A}}$, $C_{a\text{IR}}$, $C_{c\text{IR}}$ are obtained from a series of DFT static energy and phonon calculations in a similar way to what was shown above.

\begin{figure}
  \includegraphics[width=0.6\textwidth]{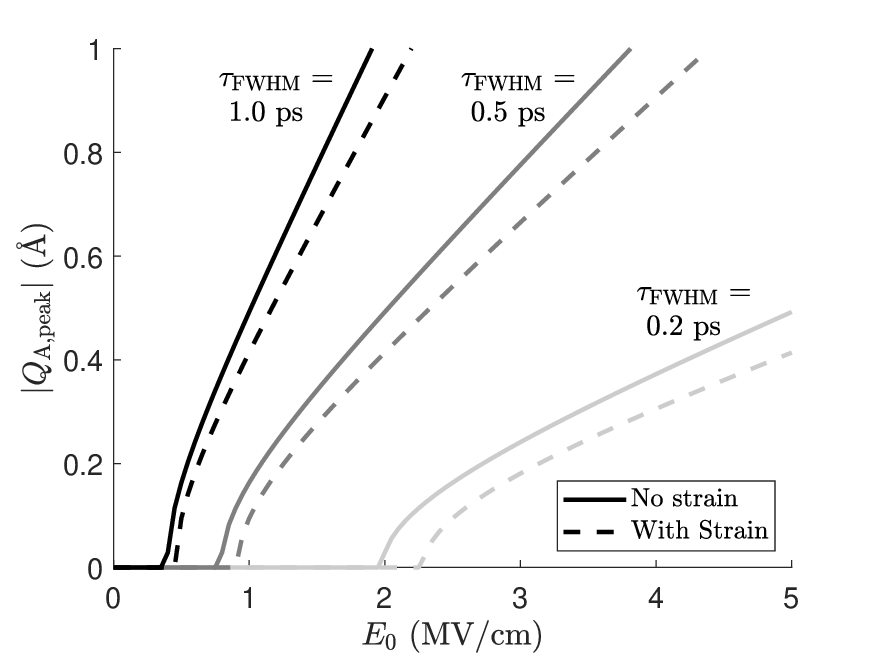}
  \caption{Peak ferroaxial mode amplitude $|Q_{\text{A,peak}}|$ vs peak electric field for light pulses with $\tau_{\text{FWHM}}$ = 0.2, 0.5, 1.0 ps, a comparison between results obtained with and without including strain.}
\label{fig:threshold_strain}
\end{figure}

With the incorporation of these additional terms due to strain, summarized in TABLE~\ref{tab:coeff_strain}, we can work out the effect of strain on the threshold electric field. Using the same procedure as in the main text, the threshold equation now contains two additional terms:
\[\langle \vert Q_{\text{A}}\vert\rangle = \text{Re} \sqrt{\frac{-K_{\text{A}}-2\left(D_{22}-\frac{C_{a\text{IR}}C_{a\text{A}}}{K_a}-\frac{C_{c\text{IR}}C_{c\text{A}}}{K_c}\right)\langle \bm{Q}_{\text{IR}}^2\rangle}{D_{\text{F}}}}\]
and $\vert Q_{\text{A,peak}}\vert = \sqrt{2}\langle\vert Q_{\text{A}}\vert\rangle$. By plotting peak ferroaxial mode amplitude $|Q_{\text{A,peak}}|$ vs peak electric field $E_0$ for light pulses with $\tau_{\text{FWHM}}$ = 0.2 ps, 0.5 ps, and 1.0 ps in FIG.~\ref{fig:threshold_strain}, we find a slightly higher threshold when strain is taken into account. 

\begin{table}[!htb]
\scriptsize
  \begin{tabular}{cc}
    \toprule \toprule
        Parameter & Value \\ \midrule
        $K_a$ & 4.17e$^5$ meV\\
        $K_c$ & 3.46e$^5$ meV\\
        $C_{a\text{A}}$ & 1.599e$^4$ meV/\AA$^2$\\
        $C_{c\text{A}}$ & 4892 meV/\AA$^2$\\
        $C_{a\text{IR}}$ & -4.89e$^5$ meV/\AA$^2$\\
        $C_{c\text{IR}}$ & -1.33e$^3$ meV/\AA$^2$\\ \bottomrule \bottomrule
        & \\
    \end{tabular}
  \caption{Fitted values of the parameters.}
  \label{tab:coeff_strain}
\end{table}

\section{IR-active modes \label{supp:IR}}
There are 11 IR-active modes in total (5 out-of-plane and 6 in-plane) excluding the acoustic modes, which correspond to zero energy translations of the crystal. We summarize their frequencies and mode-effective charges in TABLE~\ref{tab:IR}. Frequencies computed by DFT are close to those obtained by experiments performed at room temperature \cite{Wakowska_2010}. 
\begingroup
\squeezetable
\begin{table}[!htb]
	\centering
       \begin{ruledtabular}
	\begin{tabular}{ccccccc}
		Degeneracy & No. & \makecell{Computed \\ Freq (THz)} & \makecell{Computed \\ Freq (cm-1)} & \makecell{Experimental \\ Freq (cm-1)} & $\tilde{Z}^*_{x,y}$ (e) & $\tilde{Z}^*_z$ (e) \\ \cmidrule(lr){1-1}\cmidrule(lr){2-3}\cmidrule(lr){4-4}\cmidrule(lr){5-6}
		\multirow{5}{*}{1}& 2   &  3.1  & 103 & 150   &  0      &   1.841    \\ 
		& 5  &  7.8  &  259 & 256 &    0     &    1.150    \\ 
		& 7  & 10.6  &  354 & 326  &  0      &   3.133    \\ 
		& 10  &  26.1  &   870 & 921 &  0      &   5.901    \\ 
		& 11  &  28.5 &  949 & 952  &   0    &    0.112    \\ \midrule
		\multirow{6}{*}{2} & 1   &  1.7  & 56 & 98 & 0.704     &        0    \\
		& 3  &   4.1 & 137 & 188 & 0.797    &         0    \\ 
		& 4  &   6.9 & 231 & 235 & 5.550    &         0    \\ 
		& 6  &   8.6 &  288 & 302   &  1.592 &         0    \\ 
		& 8  &   10.7 &  357 & 388   & 0.762  &        0    \\
		& 9  &  23.7 & 791 & 824 & \textbf{8.610}    &         0    
	\end{tabular}
 \caption{Frequency and mode-effective charge for all the IR-active phonons. The doubly degenerate mode No. 9 with frequency = 23.7 THz has the largest mode-effective charge.\label{tab:IR}}
 \end{ruledtabular}
\end{table}
\endgroup

The mode-effective charge is defined according to Gonze et al \cite{Gonze_1997}, equation (53).  The mode-effective mass at wavevector $\bm{q}$ (here $\bm{q}=\bm{0}$) with index $m$ is calculated by
\[M^{*}_{m\bm{q}}=\sum_{\kappa\alpha}\left\{U_{m\bm{q}}(\kappa\alpha)\right\}^{2} M^{\kappa}\]
where $\alpha$ = \{$x,\ y,\ z$\}, $\kappa$ = atom index, with $U_{mq}(\kappa\alpha)$ the phonon eigendisplacements defined by equation (12) in \cite{Gonze_1997}, and normalized.
\subsection{Damping factor and its effect on response}
Wa{\'{s}}kowska et al. measured the absorbance for all IR-active modes, from which we can extract the damping factor $\gamma_{\text{IR}}$ used in the dynamical equation \cite{Wakowska_2010}. The complex dielectric function from the Lorentz dielectric model is
\[\epsilon_{\alpha\beta}(\omega)=\epsilon_{\alpha\beta}(\infty)+\frac{1}{\epsilon_0 V}\sum_n \frac{S_{n,\alpha\beta}}{\omega_{0n}^{2}-\omega^{2}-i \omega \gamma_{n}} \]
where $\alpha$ and $\beta$ label the Cartesian coordinates, V is the volume of the unit cell, $\omega_{0n}$ is the frequency of $n^{\text{th}}$ IR-active mode and $S_{n,\alpha\beta}$ is the mode-oscillator strength, as in \cite{Gonze_1997}. The index of refraction $\tilde{n}= \sqrt{\epsilon_{r}} = n+ik$ and absorbance $\alpha$ can then be expressed as \[k(\omega) \propto \sum_n \frac{\omega_{0n}^2\omega\gamma_{n}}{(\omega_{0n}^2-\omega^2)^2+\omega^2\gamma_{n}^2} \quad\alpha(\omega) = \frac{k(\omega)\omega}{c} \propto \sum_n\frac{\omega^2\gamma_{n}}{(\omega_{0n}^2-\omega^2)^2+\omega^2\gamma_{n}^2}\] 

\begin{figure}[!htb]
    \subfloat[]{\includegraphics[width=0.49\textwidth]{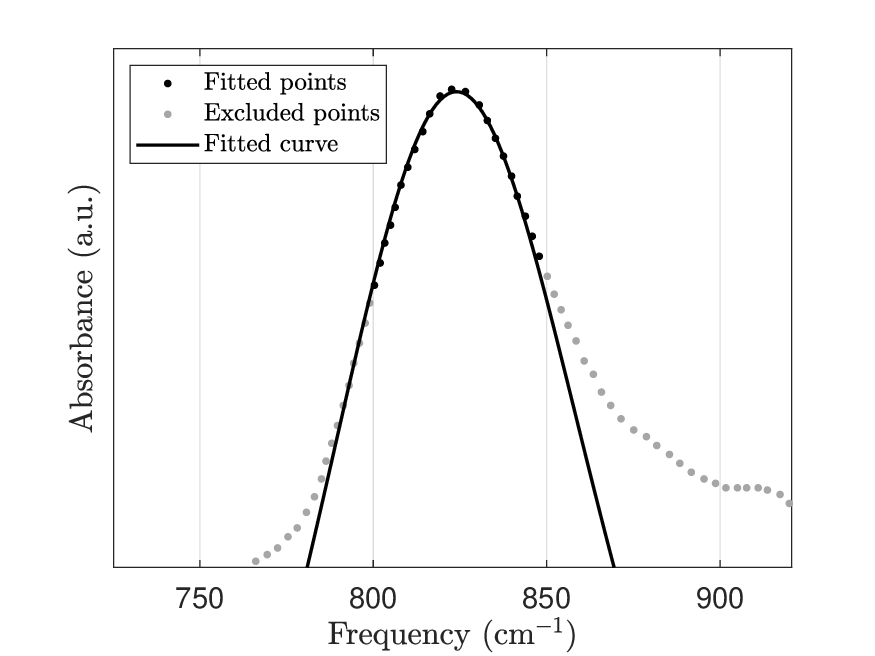}} \hfill
    \subfloat[]{\includegraphics[width=0.49\textwidth]{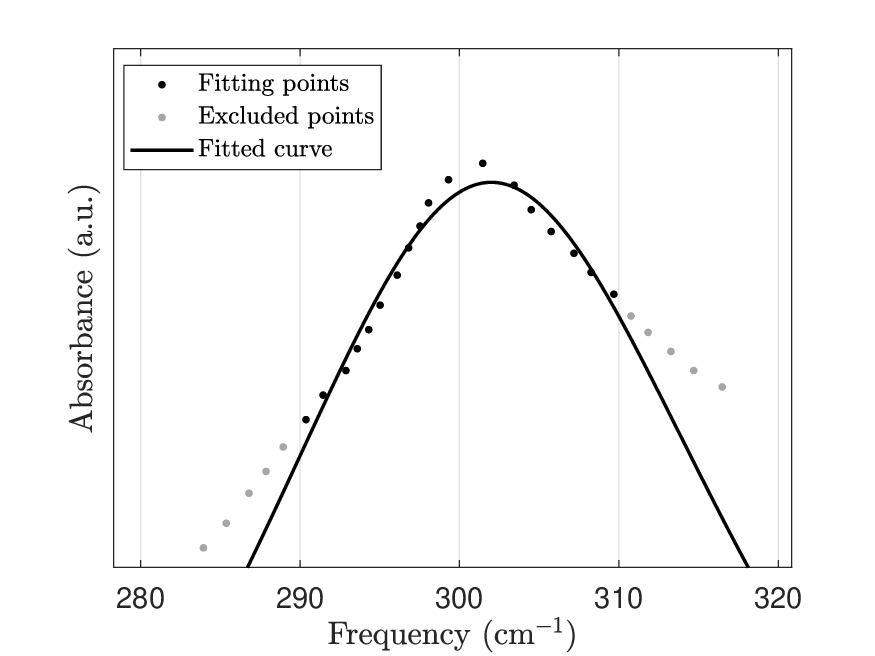}}
    \caption{Fit to the experimental absorbance data used for calculating the damping. Data extracted from Ref. \cite{Wakowska_2010}}
    \label{fig:absorb}
\end{figure}

Here we are interested in the peak corresponding to IR No. 9 for light-induced ferroaxial order, or IR No. 6 in the case of switching. We fit the absorbance data around the two peaks using $\omega_{0}$ = 824 cm$^{-1}$, and $\omega_{0}$ = 302 cm$^{-1}$ respectively, from the experimental data (FIG.~\ref{fig:absorb}). The fitting gives $\gamma_{\text{IR}}$ = 119.4 cm$^{-1}$ $\approx$ 4 THz for IR No. 9 and $\gamma_{\text{IR}}$ = 41 cm$^{-1}$ $\approx$ 1.3 THz for IR No. 6.

\begin{figure}[!htb]
    \centering
    \includegraphics[width=0.8\textwidth]{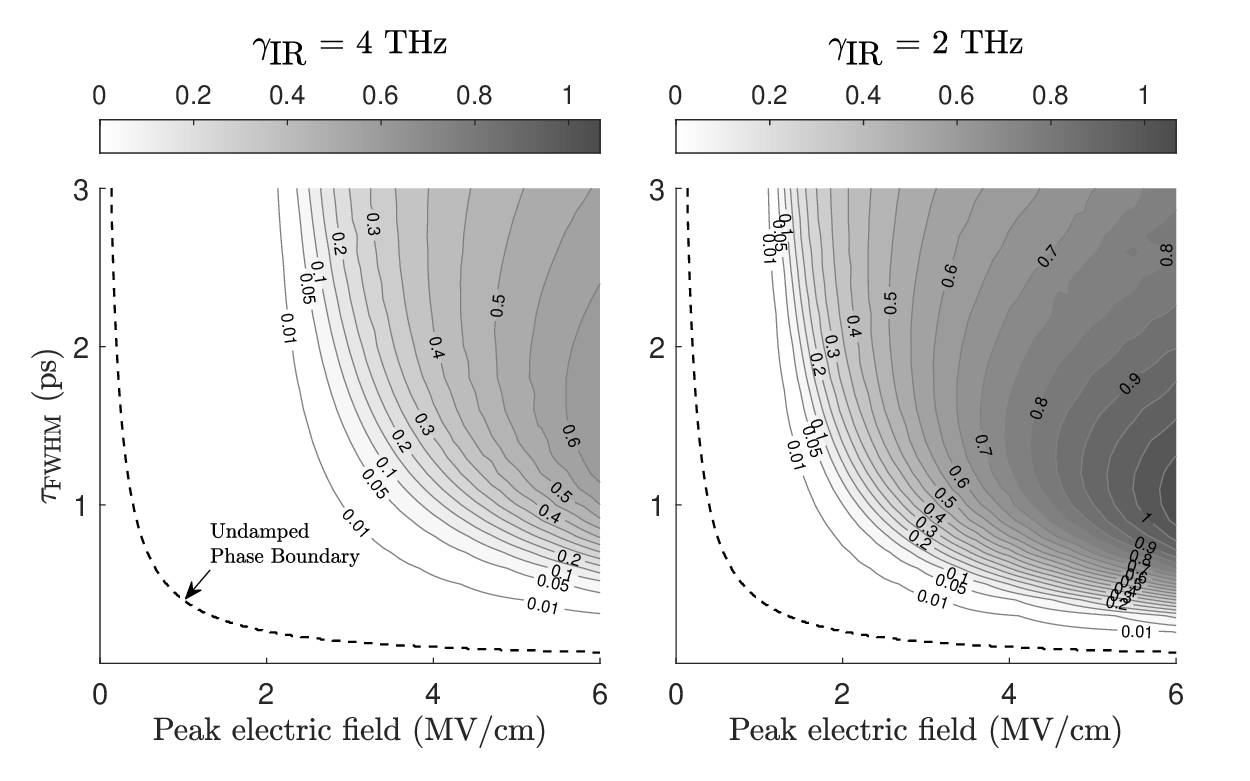}
    \caption{Contour plot of peak ferroaxial amplitude with $\gamma_{\text{IR}}$ = 2 THz and 4 THz. }
    \label{fig:QFA_2THz}
\end{figure}

The threshold between trivial response and large growth of the ferroaxial mode is strongly sensitive to damping. FIG.~\ref{fig:QFA_2THz} shows the phase diagram for $\gamma_{\text{IR}} = 2 $ THz and 4 THz. A side-by-side comparison clearly indicates that with weaker IR-active phonon damping, the peak ferroaxial amplitude increases, and the numerical phase boundary moves closer to the undamped analytical phase boundary (dashed curve).

\subsection{Structural insight into $D_{22}$}
\begin{figure}[!htb]
    \subfloat[No. 9 direction 1]{\includegraphics[width=0.4\textwidth]{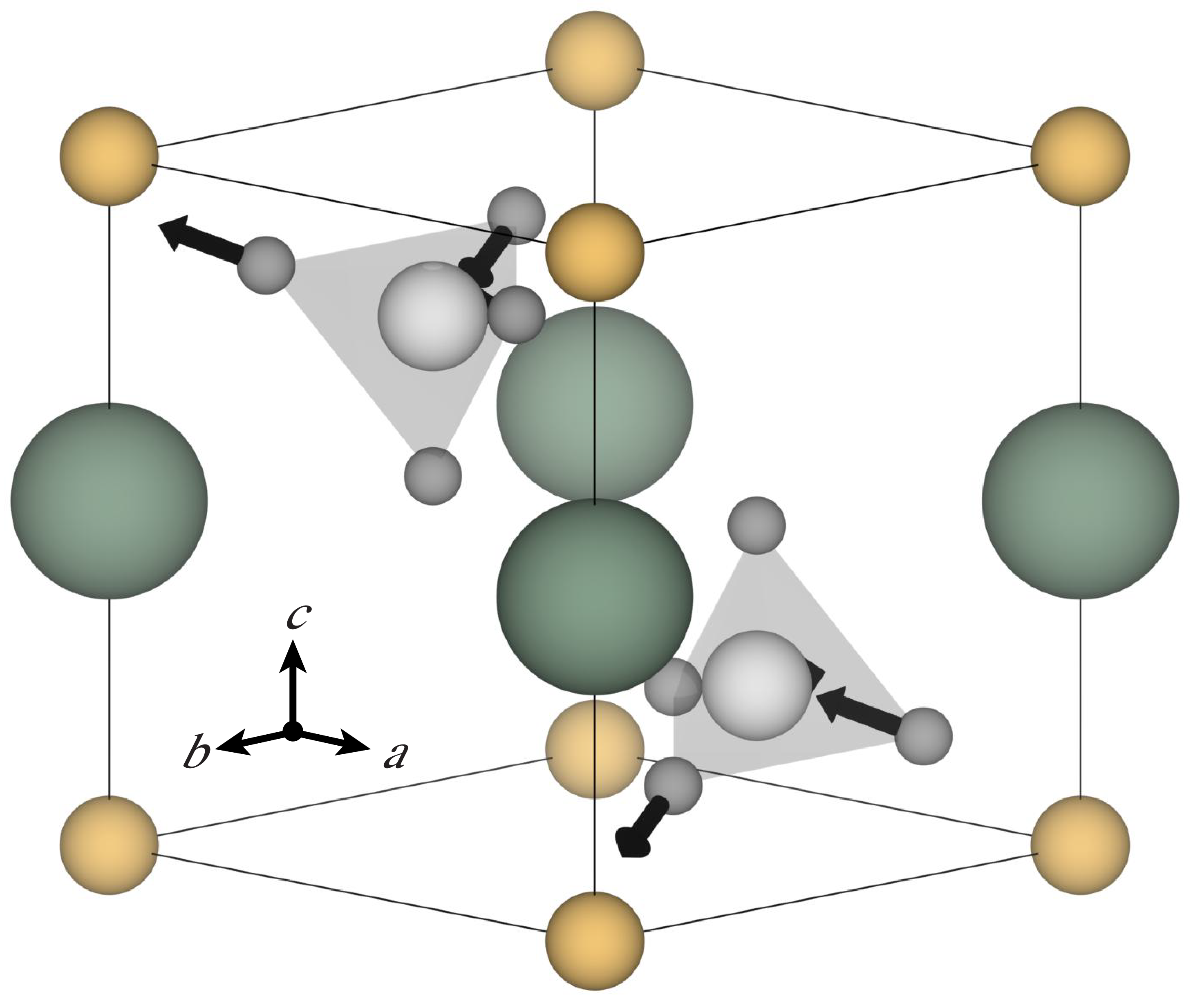}} \hfill
    \subfloat[No. 9 direction 2]{\includegraphics[width=0.4\textwidth]{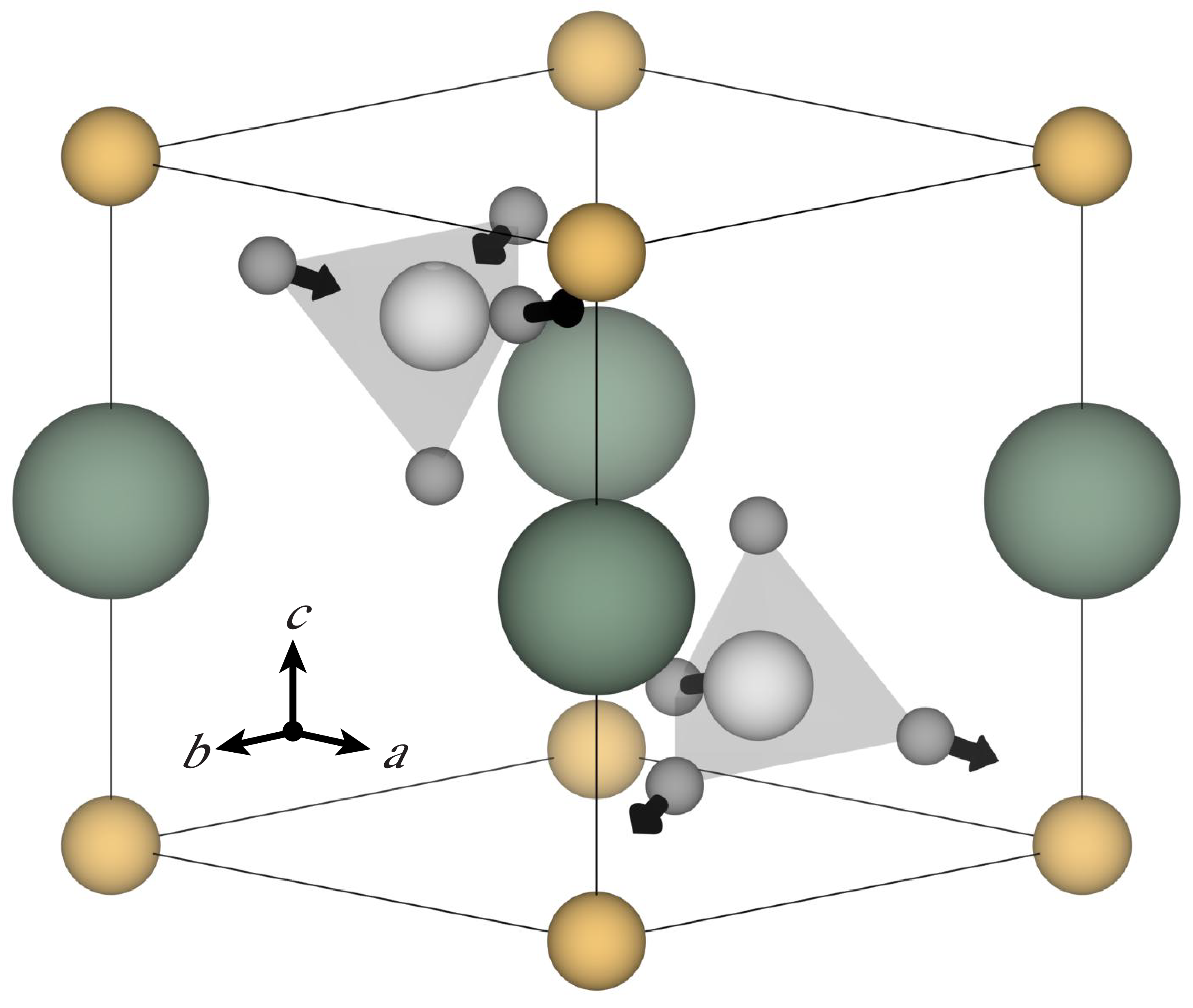}}
    
    \subfloat[No. 6 direction 1]{\includegraphics[width=0.4\textwidth]{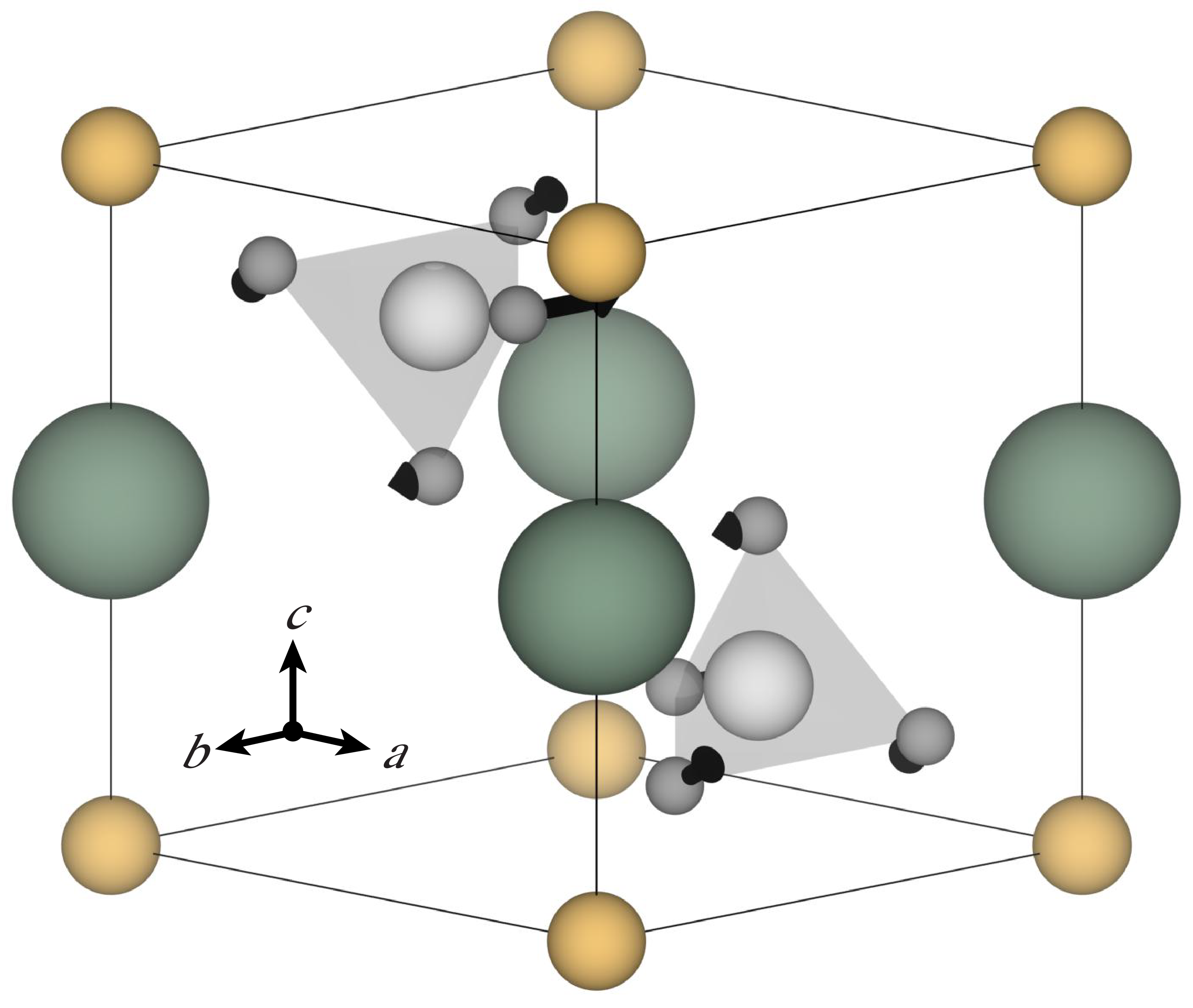}} \hfill
    \subfloat[No. 6 direction 2]{\includegraphics[width=0.4\textwidth]{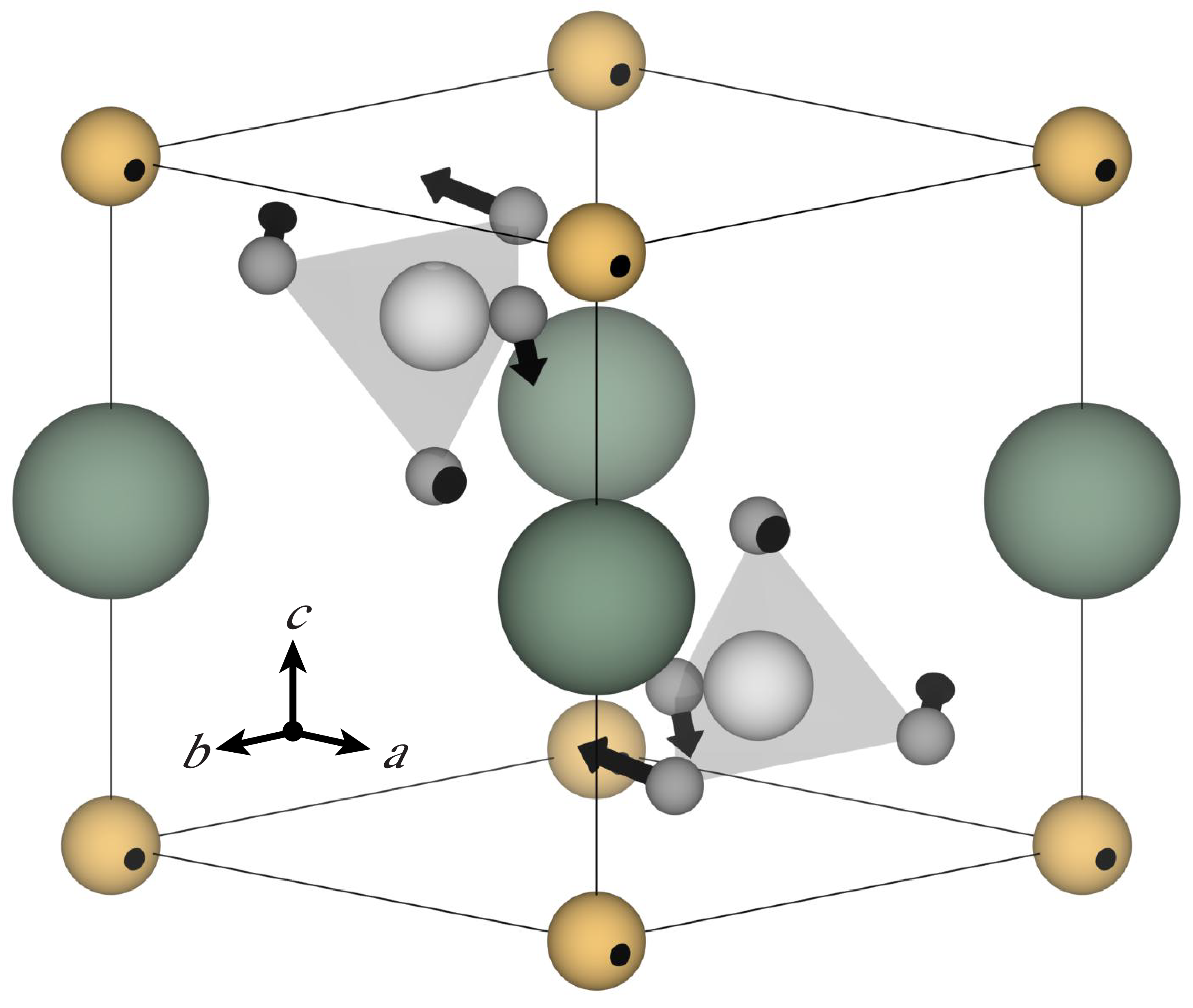}}
    \caption{Phonon eigen-displacement for doubly degenerate IR-active modes No. 9 at 23.7 THz and No. 6 at 8.6 THz.}
    \label{fig:eigendisp}
\end{figure}

In the main text, we use a doubly degenerate mode (No. 9) which has the largest negative $D_{22}$ to induce ferroaxial order. Visualization of this mode shows mainly displacement of O, with a small contribution of Mo (FIG.~\ref{fig:eigendisp}). The large negative $D_{22}$ can be explained by a geometric argument. The movements of oxygens change both the Mo-O and Fe-O bond lengths. When the IR-active mode amplitude increases, some of the Mo-O and Fe-O bonds will become shorter, which is unfavorable energetically. The ferroaxial mode, which rotates the tetrahedron, can lower the energy by lengthening these shortened bonds.

Mode No. 6, which is used for ferroaxial domain switching, involves almost purely oxygen displacements (FIG.~\ref{fig:eigendisp}). It has slightly positive $D_{22}$, which can be understood by the increased distortion of the MoO$_4$ tetrahedra when both the IR-active mode and the ferroaxial mode are concurrently displaced.

\subsection{Effect of the IR-active phonon on other modes}
\begin{figure}[!htb]
    \subfloat[direction 1]{\includegraphics[width=0.49\textwidth]{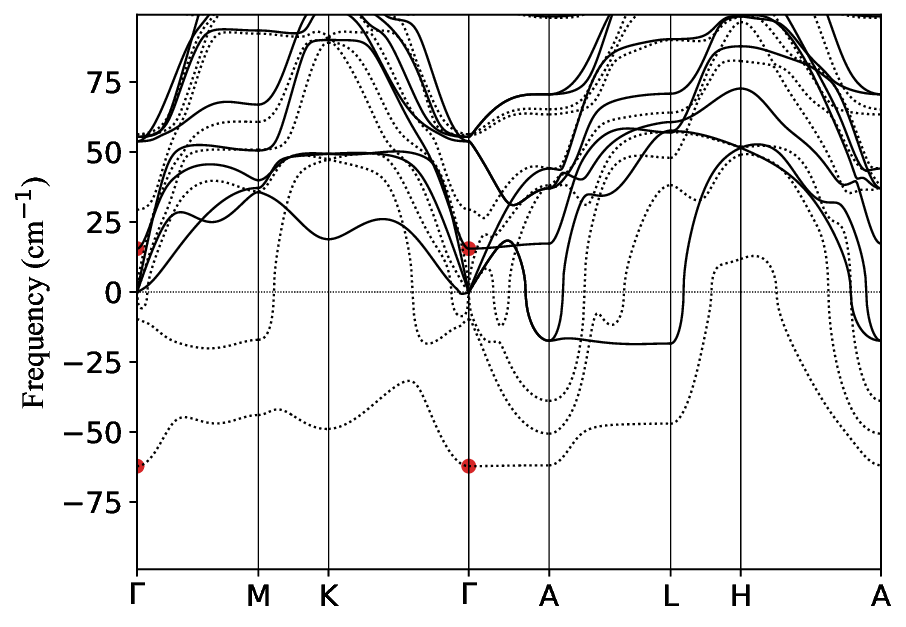}} \hfill
    \subfloat[direction 2]{\includegraphics[width=0.49\textwidth]{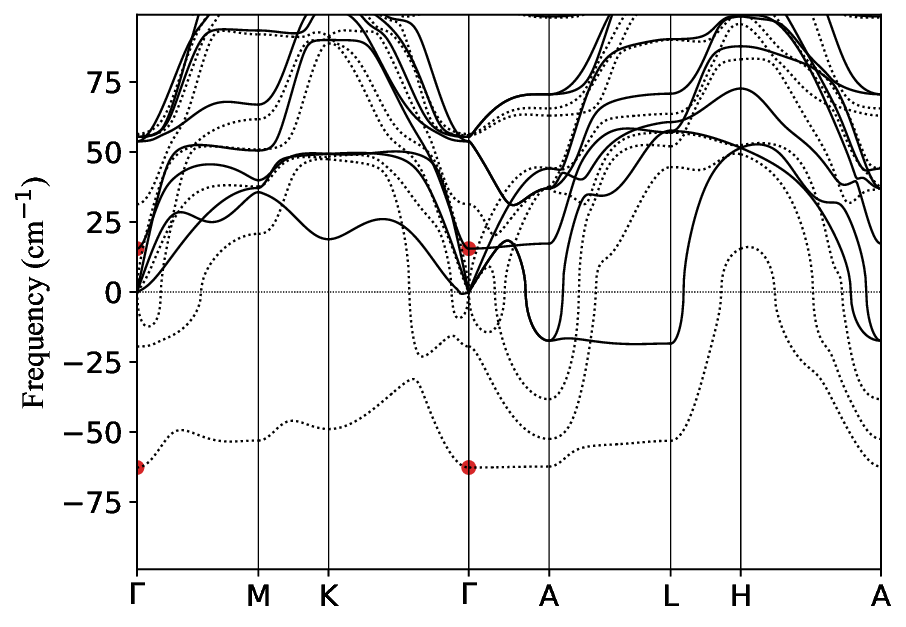}}
    \caption{A comparison between phonon dispersion of the fully relaxed parent structure (solid curves) and with IR-active mode No. 9 of amplitude = 0.1 \AA \ added into the structure (dotted curves). The red circles show a decrease in ferroaxial phonon frequency in the presence of IR No. 9.} \label{fig:phonon_IR}
\end{figure}

The excitation of IR-active phonon may affect other phonon degrees of freedom besides the ferroaxial phonon. The effect is shown in Fig.~\ref{fig:phonon_IR} for IR No. 9 in both directions.

\section{Robustness of results \label{supp:robust}}
\subsection{Unidirectionality of the ferroaxial mode as a function of the NLP strength $C_c$}
\begin{figure}[!htb]
    \subfloat[]{\includegraphics[width=0.49\textwidth]{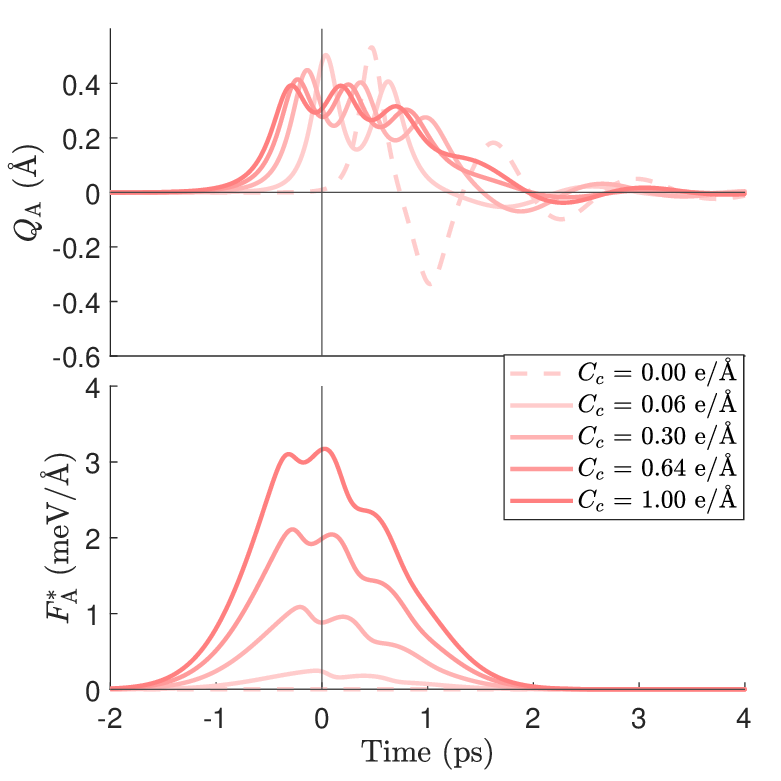}} \hfill
    \subfloat[]{\includegraphics[width=0.49\textwidth]{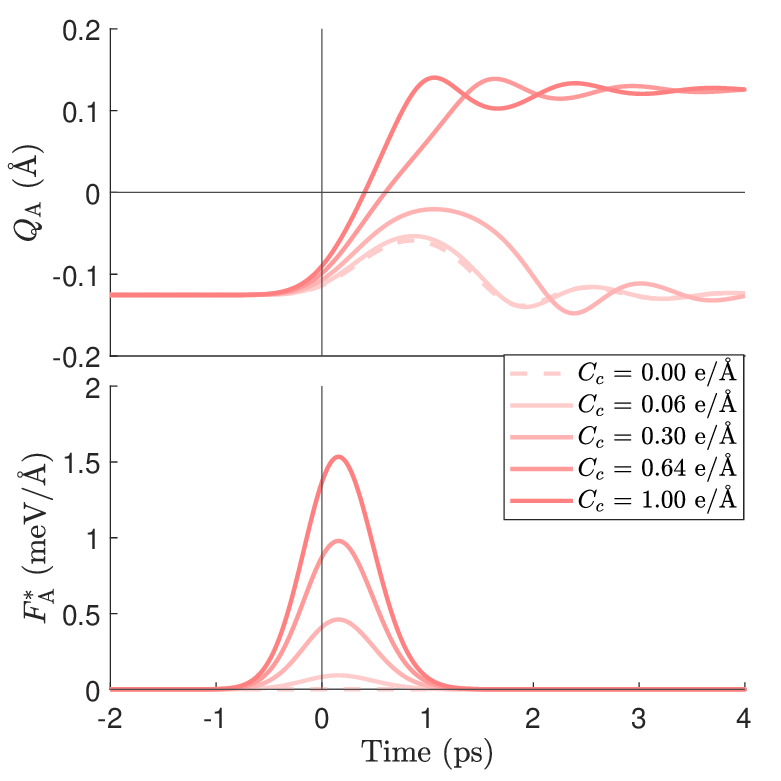}}
    \caption{Dynamical simulation with left-circularly polarized light. $C_c$ is varied, but all other parameters are fixed. (a) Inducing ferroaxial order using light at 23.7 THz, $E_0$ = 5 MV/cm and $\tau_{\text{FWHM}}$ = 2 ps. (b) Switching ferroaxial domain using light at 8.6 THz, $E_0$ = 3 MV/cm and $\tau_{\text{FWHM}}$ = 1 ps.}
    \label{fig:dynamics_Cc}
\end{figure}

The size of the NLP parameter $C_c$ is varied in this section to show that decreasing the value found by DFT for RbFe(MoO$_4$)$_2$ by an order of magnitude preserves the directional control of induced ferroaxial order. The simulated dynamics are shown in FIG.~\ref{fig:dynamics_Cc}a where even a small $C_c$ allows for unidirectional control. Though smaller $C_c$ may forbid switching, shown in FIG.~\ref{fig:dynamics_Cc}b, we can always increase the light intensity or tune the system closer to the phase boundary to make switching happen.

\subsection{Insensitivity to initial condition}
\begin{figure}[!htb]
    \centering
    \includegraphics[width=\textwidth]{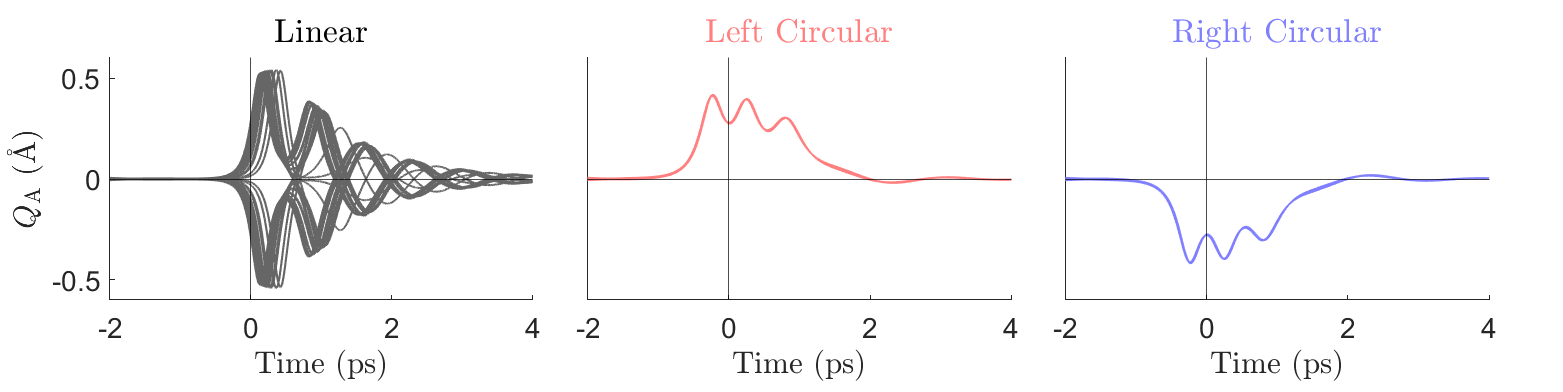}
    \caption{Simulation starting from different initial conditions. Light pulse with $E_0$ = 5 MV/cm and $\tau_{\text{FWHM}}$ = 2 ps is used.}
    \label{fig:init_condition}
\end{figure}

Due to vibrations at finite temperature, the ferroaxial mode may be slightly perturbed when the pulse excites the mode. Therefore, the effect of initial conditions on the response needs to be investigated. The ferroaxial mode is initialized with different small displacements and velocities. In FIG.~\ref{fig:init_condition}, the response to linearly polarized light shows variations, but the response to circularly polarized light is insensitive to the initial conditions. 

\subsection{Response to different light pulses}
\begin{figure}[!htb]
    \centering
    \includegraphics[width=\textwidth]{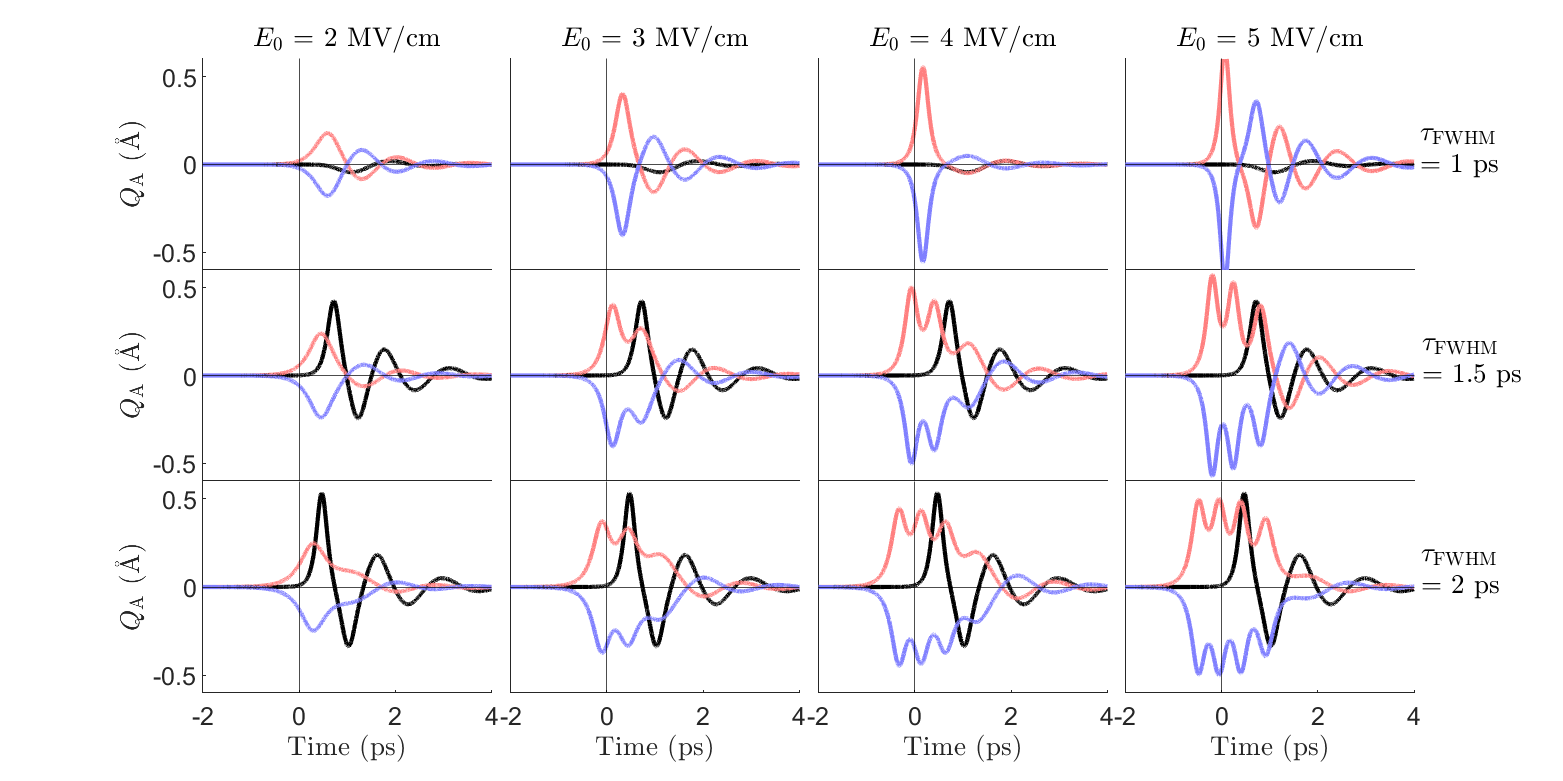}
    \caption{Response of ferroaxial mode to pulses of various peak electric fields $E_0$ and durations $\tau_{\text{FWHM}}$. Linear, left- and right-circular polarizations are indicated by black, red and blue colors.}
    \label{fig:robust}
\end{figure}

Our simulations in FIG.~\ref{fig:robust} show that domain selectivity and transient unidirectional responses are observed for all pulses with intensity above the threshold.

\section{Switching criteria \label{supp:switch}}
\begin{figure}[!htb]
    \subfloat[]{\includegraphics[width=0.49\textwidth]{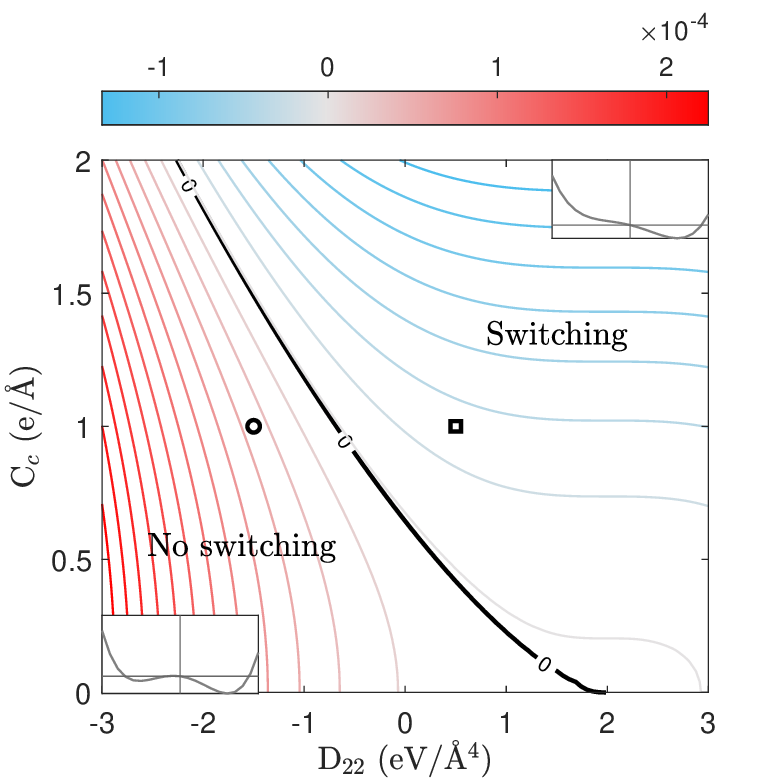}} \hfill
    \subfloat[]{\includegraphics[width=0.49\textwidth]{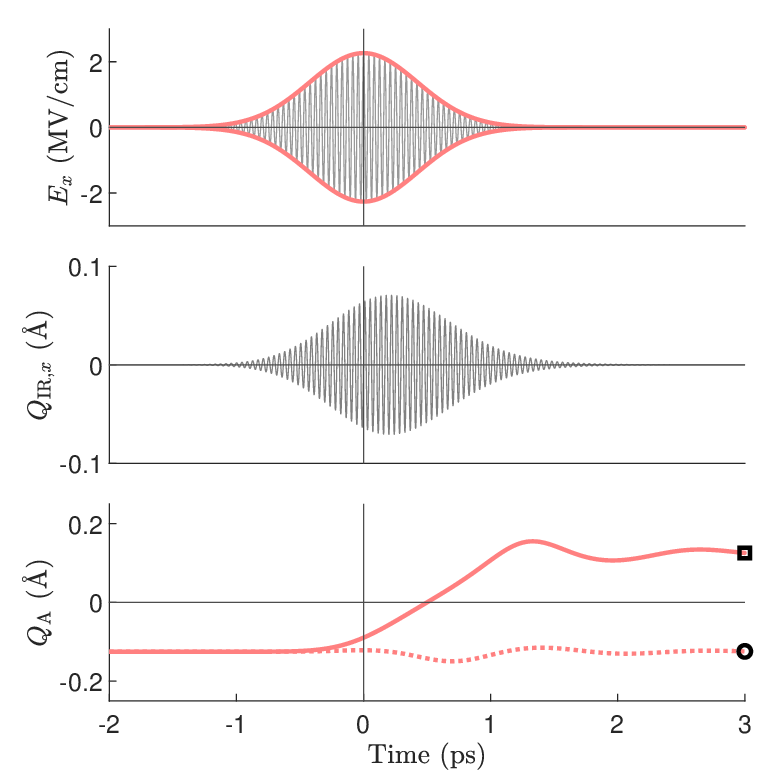}}
    \caption{(a) Contour plot of $\Delta$ versus $D_{22}$ and $C_c$. Switching occurs in the blue region when $\Delta < 0$. The parameters indicated by the open square (switching) and open circle (no switching) are used for dynamical simulation. (b) Dynamical simulation for a light pulse of E$_\text{peak}$ = 3.2 MV/cm and $\tau_{\text{FWHM}}$ = 1 ps at 23.7 THz with left-circular polarization, which generates $\vert\bm{Q}_{\text{IR,peak}}\vert\approx$ 0.1 \AA.}
    \label{fig:switch_IR9}
\end{figure}

The local minima on the free energy are found by taking the derivative with respect to $Q_{\text{A}}$
\[
	D_{\text{A}}\langle Q_{\text{A}}^3 \rangle + \left(K_{\text{A}}+2D_{22} \langle\bm{Q}_{\text{IR}}^2 \rangle\right)\langle Q_{\text{A}}\rangle + C_c( Q_{\text{IR},x}E_y-Q_{\text{IR},y}E_x) = 0 .
\]
In the ideal situation where $Q_{\text{IR},x}$ and $E_y$, and similarly, $Q_{\text{IR},y}$ and $E_x$, are in-phase, and they reach the maximum simultaneously, we have
\[
	D_{\text{A}}\langle Q_{\text{A}}^3 \rangle + \left(K_{\text{A}}+2D_{22} \langle\bm{Q}_{\text{IR}}^2\rangle \right)\langle Q_{\text{A}}\rangle \pm\frac{1}{2} C_c \vert\bm{Q}_{\text{IR,peak}}\vert E_0 = 0 .
\]
This has the form $Q_{\text{A}}^3+pQ_{\text{A}}+q=0$ where 
\[p = \frac{K_\text{A}+2D_{22}\langle\bm{Q}_{\text{IR}}^2\rangle}{D_{\text{A}}}\quad \text{and}\quad q=\frac{C_c\vert\bm{Q}_{\text{IR,peak}}\vert E_0}{2D_{\text{A}}}\]
For there to be no switching barrier, which is equivalent to when the cubic equation has 1 real root and 2 imaginary roots, requires
\[
	\Delta \equiv -4p^3-27q^2 < 0
\]
In FIG.~\ref{fig:switch_IR9}a, we plot $\Delta$ as a function of $D_{22}$ and $C_c$, for a representative case where $\bm{Q}_{\text{IR,peak}}=0.1$ \AA\ for the 23.7 THz mode. The solid black line separates the region of parameter space where switching of the ferroaxial domain is possible (blue contours) from where it is not possible (red contours). Selected dynamics are shown for parameters drawn from FIG.~\ref{fig:switch_IR9}a to illustrate domain switching evolution in the two regions of parameter space. FIG.~\ref{fig:switch_IR9}b shows the circularly-polarized light field, IR-active phonon, and ferroaxial mode dynamics versus time for the two points highlighted in FIG.~\ref{fig:switch_IR9}a, represented by the open circle and square. The dynamics show agreement with the approximations made for switching. Since the 23.7 THz mode has $D_{22}$ = -7.676 eV/\AA$^4$ and $C_c$ = 0.638 e/\AA, which sits in the ``No switching" region, our simulations suggest that exciting the 23.7 THz mode will not lead to energy barrier lowering and switching.

Our simulation of switching in the main text chooses the coefficient $K_\text{A}$ to be slightly negative, corresponding to a condition just below the critical temperature. Here, we discuss the temperature range within which the switching can be effective. Let the coefficient $K_\text{A}$ be temperature dependent, i.e. $K_\text{A}=-a\vert\Delta T\vert$, where $a$ is a positive, material-dependent constant and $\vert\Delta T\vert$ is the amount of temperature below the critical point. For the previously discussed switching criterion to be fulfilled \[p>(-\frac{27}{4}q^2)^{1/3}\] or
\[\vert\Delta T\vert < \frac{1}{a}\left[ 2D_{22}\langle\bm{Q}_{\text{IR}}^2\rangle+3\left(\frac{1}{4}D_{\text{A}}C_c^2\vert\bm{Q}_{\text{IR,peak}}\vert^2 E_0^2\right)^{1/3}\right]\]
This expression is most sensitive to the $D_{22}$ and $a$. With a linear dependence of $Q_{\text{IR}}$ on $E_0$, $\left| \Delta T \right|$ is expected to be dominated by quadratic dependence on the peak electric field through the $D_{22}$ parameter with a weaker (i.e. $E_0^{4/3}$) dependence on $C_c$ and $D_\text{A}$.

\section{X-ray Measurement \label{supp:measure}}
\begin{figure}[!htb]
\centering
\includegraphics[width=0.6\textwidth]{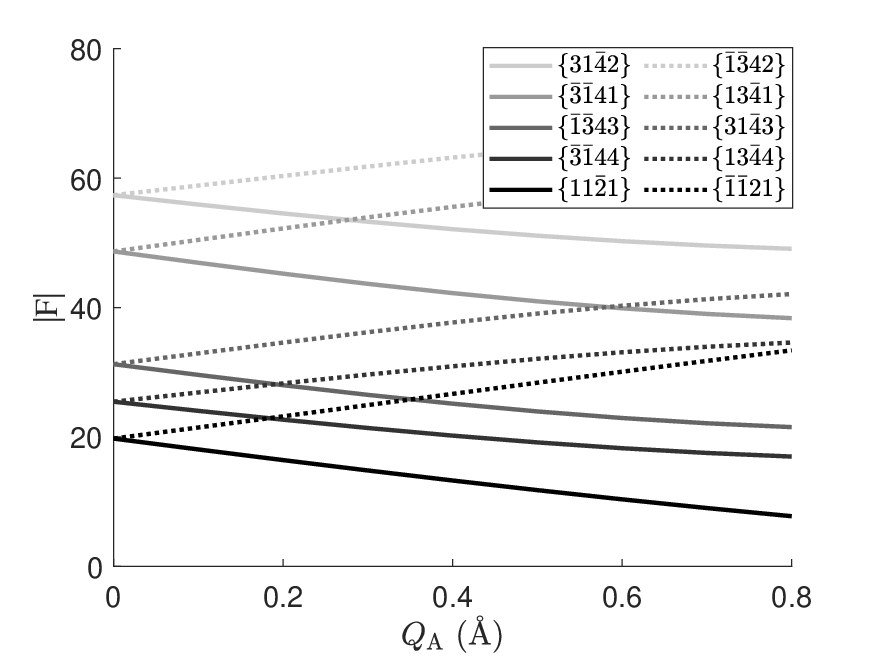}
    \caption{Norm of structure factors v.s. ferroaxial mode amplitude. The data is computed by VESTA, taking the X-ray (anomalous) dispersion into account.}
    \label{fig:SF}
\end{figure}

X-ray diffraction has been used to characterize single crystal RbFe(MoO$_4$)$_2$. The ferroaxial phase transition lowers the space group such that some of the degeneracies of structure factors in the high-symmetry P$\Bar{3}$m1 phase are lifted in the low-symmetry P$\Bar{3}$ phase. We select a few structure factors that have the largest splitting upon the symmetry lowering and plot how the norms of structure factors change as we increase the ferroaxial mode amplitude in FIG.~\ref{fig:SF}.

\bibliography{ref}